\documentclass[usenatbib,useAMS]{mn2e} 
\usepackage{times}
\usepackage{epsf}

\newcommand\jnlstyle{\rmfamily}
\newcommand\refjnl[1]{{\jnlstyle#1}}
\newcommand\aj{\refjnl{AJ}}

\newcommand\apj{\refjnl{ApJ}}
          
\newcommand\apjl{\refjnl{ApJ}}
          
\newcommand\apjs{\refjnl{ApJS}}

\newcommand\mnras{\refjnl{MNRAS}}

\newcommand\pasp{\refjnl{PASP}}

\newcommand\nat{\refjnl{Nature}}

\newcommand\physrep{\refjnl{Phys.~Rep.}}

\newcommand{\mm}[1]{\mbox{$#1$}}
\newcommand{\unit}[1]{\ifmmode \:\mbox{\rm #1}\else \mbox{#1}\fi}
\newcommand{\bvr}[1]{\bmath{#1}} 
\newcommand{\sbr}[1]{_{\mathrm{#1}}} 
\newcommand{\expec}[1]{\mm{\left\langle #1 \right\rangle}} 
\newcommand{\mone}{\mm{^{-1}}}

\newcommand{\ten}[1]{\mm{\times 10^{#1}}}

\newcommand{\kms}{\unit{km~s\mone}} 
\newcommand{\kpc}{\unit{kpc}} 
\newcommand{\mpc}{\unit{Mpc}}  
\newcommand{\hkpc}{\mm{h\mone}\kpc} 
\newcommand{\hmpc}{\mm{h\mone}\mpc}

\newcommand{\lb}[2]{\mm{l = #1\degr}, \mm{b = #2\degr}} 
\newcommand{\lberr}[4]{\mm{l = #1\degr\pm#2\degr}, \mm{b = #3\degr\pm#4\degr}} 
 
\newcommand{\lbapr}[2]{\mm{l \approx #1\degr}, \mm{b \approx #2\degr}}

\newcommand{\wrt}{with respect to}

\newcommand{\mg}{\mbox{Mg$_{2}$}} 
 
\newcommand{\dnsig}{\mm{D\sbr{n}-\sigma}} 
\newcommand{\om}{\mm{\Omega\sbr{m}}} 
\newcommand{\omsix}{\mm{\om^{0.6}}}

\newcommand{\rhat}{\mm{\hat{\bvr{r}}}} 
\newcommand{\sig}{\mbox{$\sigma$}}

\newcommand{\bvmg}{\mbox{$(B-V)$} -- \mg}

\newcommand{\re}{\mbox{$R\sbr{e}$}}
\newcommand{\sbe}{\mbox{$\expec{\mu}\sbr{e}$}}

\newcommand{\secref}[1]{Section~\ref{sec:#1}} 
 
\newcommand{\eqref}[1]{equation~(\ref{eq:#1})} 
\newcommand{\Eqref}[1]{Equation~(\ref{eq:#1})} 
\newcommand{\figref}[1]{Fig.~\ref{fig:#1}} 
\newcommand{\tabref}[1]{Table~\ref{tab:#1}} 

\newcommand{\vext}{\mbox{$\bvr{V}\sbr{ext}$}}

\title
[Streaming motions of galaxy clusters --- V. The peculiar velocity field]
{Streaming~motions~of~galaxy~clusters~within~12000\,\kms~---  \\
V. The~peculiar~velocity field} 
\author [M.J. Hudson et al.] 
{
Michael J. Hudson$^1$,
Russell J. Smith$^1$,
John R. Lucey$^2$
and
Enzo Branchini$^3$
\\ 
$^1$ Department of Physics, University of Waterloo,
Waterloo, Ontario N2L 3G1, Canada.\\
E-mail: mjhudson@uwaterloo.ca,rjsmith@astro.uwaterloo.ca\\
$^2$ Department of Physics, University of Durham, South Road, Durham
DH1 3LE, United Kingdom. E-mail: John.Lucey@durham.ac.uk\\
$^3$ Dipartimento di Fisica, Universit\`{a} degli Studi Roma TRE, Roma, Italia
}

\begin{document} 

\maketitle 

\begin{abstract} 
  
  We analyze in detail the peculiar velocity field traced by 56
  clusters within 120 \hmpc\ in the ``Streaming Motions of Abell
  Clusters'' (SMAC) sample.  The bulk flow of the SMAC sample is
  $687\pm203$ \kms, toward \lberr{260}{13}{0}{11}. We discuss possible
  systematic errors and show that no systematic effect is larger than
  half of the random error.  The flow does not drop off significantly
  with depth, which suggests that it is generated by structures on
  large scales.  In particular, a Great Attractor as originally
  proposed by Lynden-Bell et al. cannot be responsible for the SMAC
  bulk flow. The SMAC data suggest infall into an attractor at the
  location of the Shapley Concentration, but the detection is marginal
  (at the 90\% confidence level).  We find that distant attractors in
  addition to the Shapley Concentration are required to explain the
  SMAC bulk flow.  A comparison with peculiar velocities predicted
  from the \emph{IRAS} PSCz redshift survey shows good agreement with
  a best fit value of $\beta\sbr{I} = \omsix/b\sbr{I} = 0.39\pm0.17$.
  However, the PSCz density field is not sufficient to acount for all
  of the SMAC bulk motion. We also detect, at the 98\% confidence
  level, a residual bulk flow of $372\pm127\,\kms$ toward \lb{273}{6}
  which must be generated by sources not included in the PSCz
  catalogue, that is, either beyond 200 \hmpc, in the Zone of
  Avoidance or in superclusters undersampled by \emph{IRAS}. Finally,
  we compare the SMAC bulk flow with other recent measurements. We
  argue that, at depths ranging from 60 to 120 \hmpc, flows of order
  600 \kms\ are excluded by multiple data sets.  However, convergence
  to the CMB frame by a depth of 60 \hmpc\ is also excluded by
  multiple data sets.  We suggest that a bulk flow of $225\,\kms$
  toward \lb{300}{10} at depths greater than 60 \hmpc\ is consistent
  with all peculiar velocity surveys, when allowance is made for
  sparse sampling effects.
\end{abstract}


\section{Introduction} 
\label{sec:intro} 

The source of the Local Group's (LG) peculiar velocity of $627\pm22
\kms$ toward \lb{276}{30} \citep{KogLinSmo93} with respect to the
Cosmic Microwave Background (CMB) has been a puzzle since the
detection of the CMB dipole \citep{SmoGorMul77}.  In order to resolve
this fundamental question, it is necessary to map the peculiar
velocities of nearby galaxies.  If a volume which is at rest with
respect to the CMB can be identified, then the masses responsible for
the motion of the LG must be contained within that volume.

An important milestone in peculiar velocity surveys was the study of
\citet{LynFabBur88}, who used the \dnsig\ distance indicator and
claimed detection of a ``Great Attractor'' (GA) at a distance of
approximately 45 \hmpc\ believed to be responsible for most the LG's
motion.  Subsequent peculiar velocity surveys of the nearby Universe
undertaken in the early 90s did not reveal the expected infall
signature on the far side of the GA \citep{MatForBuc92} and hinted at
a large coherence length for the flow \citep{Wil90,Cou92}. Analyses of
redshift surveys \citep{Dre88,StrDavYah92,Hud93,Hud94b} did identify a
significant excess of galaxies near the proposed GA, but with an
abundance insufficient to generate the LG's motion.  For example,
\citet{Hud94b} concluded that $400\pm45\,\kms$ of the LG's motion
arose from sources beyond 80 \hmpc.  The location of the peak of the
GA in the Zone of Avoidance (ZOA), as given by \citet{KolDekLah95},
suggests the possibility that most of the GA's overdensity may be
obscured. Searches in the ZOA have revealed the presence of massive
clusters such as A3627 \citep{KraWouCay96} but the integrated
overdensity within the ZOA appears still insufficient to generate
substantial infall at the LG \citep{StaJurHen00}.

\citet[hereafter LP]{LauPos94} used the photometric properties of
brightest cluster galaxies as standard candles and claimed that a much
larger region of space extending to 150 \hmpc\ was moving at a
velocity of $689\pm178\,\kms$ toward \lb{343}{52}.  The LP result was
puzzling, first because the amplitude of the flow on such a large
scale was higher than expected in popular cosmological models
\citep{StrCenOst95}, and second because the direction of the LP flow
was significantly different from previous bulk flow measurements
\citep{StrWil95}.

Recent surveys of field galaxies to depths $R \la 60\,\hmpc$
generally suggest that the bulk flow within these nearby volumes is in
the range from $100$ -- $300\,\kms$ in the direction \lbapr{300}{10}
\citep{WilCouFab97, GioHayFre98, CouWilStr00,
daCBerAlo00b,TonBlaAjh00}. Clusters surveys on large ($R \ga 100
\hmpc$) scales have not supported the LP result, but, at face value,
also appear to suggest a wide range of values for the large-scale
motion \citep{MouAkeBot93,HudSmiLuc99, Wil99b, DalGioHay99b,
ColSagBur01}.  This apparent disagreement is due to neglect of sparse
sampling on the quoted flow errors \citep{WatFel95}. When the effects
of sparse sampling are properly taken into account, the large-scale
surveys are not in conflict \citep{Hud03}.  For recent reviews, the
reader is directed to \citet{CouDek01} and \citet{Zar02b}.

This paper is the fifth in a series based on the ``Streaming Motions
of Abell Clusters'' (SMAC) project. The aim of this project is to
obtain distance estimates, via the Fundamental Plane (FP) method, for
elliptical galaxies in clusters and to map the peculiar velocity field
within 120 \hmpc\ of the Local Group.  First results from this work
were presented in \citet{HudSmiLuc99}, who quoted a bulk flow of
$630\pm200\,\kms$ toward \lb{260}{-1} from a preliminary analysis of
the same sample studied here.

Previous papers in this series have reported new data for this
project. \citet[hereafter SMAC-I and SMAC-II
respectively]{SmiLucHud00, SmiLucSch01} presented spectroscopic and
photometric data, respectively; these data were compared and combined
with other data available in the literature to obtain the final data
set used for this project \citep[hereafter SMAC-III]{HudLucSmi01}.
Smith et al. (in preparation, hereafter SMAC-IV) reports FP distances
for 56 clusters.

In this paper, we analyze the peculiar velocity field in the Universe
as traced by the SMAC clusters.  In \secref{data}, we summarize the
SMAC peculiar velocity data. In \secref{bulk}, we model the peculiar
velocity field as a simple bulk flow (the dipole moment of the
velocity field).  An important aspect of this paper is to evaluate the
robustness of the results to systematic effects (\secref{sys}).  In
\secref{othermod} we consider more complicated flow models, including
toy models based on simple attractors.  \secref{pscz} compares the
peculiar velocities of SMAC clusters to the predictions from the
\emph{IRAS} PSCz density field.  \secref{comparison} compares the flow
of the SMAC sample to other results in the literature.

\section{Data}
\label{sec:data}

In SMAC-IV, we tabulated peculiar velocity data for 56 clusters
spanning the whole sky, and extending to a distance of $\sim 120
\hmpc$.  The peculiar velocities quoted there were based on a inverse
fit to the FP, in which $\log \sig$ is regressed on $\log \re$ and
\sbe.  The inverse fit has the advantage that it is insensitive to
selection on the photometric parameters.

Median peculiar velocity errors per cluster are $\sim 575\kms$.  These
errors are primarily due to intrinsic scatter in the Fundamental Plane
which contributes a fractional distance error of $0.21/\sqrt{N}$,
where $N$ is the number of galaxies in the cluster.  However, the
errors also include the effect of the 16\% uncertainties in the
\citet[hereafter SFD]{SchFinDav98} extinctions and uncertainties in
the mean redshifts of the clusters.

In SMAC-IV, we also quoted for each cluster the error contribution
arising from uncertainties in matching velocity dispersions from
disparate observing runs.  An important aspect of this work is the
treatment of these errors, which are not independent from cluster to
cluster, but are correlated on the sky.  To calculate their effect on
the bulk flow, we bootstrap re-sample the ``overlap'' sample used to
calculate the velocity dispersion matching corrections.  We then
generate new FP datasets using those bootstrapped corrections and use
these bootstrap samples to calculate the covariance matrix of the
cluster peculiar velocities. We find that these systematic errors do
not dominate the error budget in the bulk flow. On the other hand, due
to their coherent nature, neither can they be neglected (see
\secref{specsys} below).

\section{The Bulk Flow}
\label{sec:bulk}

\subsection{Flow Model}

One statistic of particular interest is the bulk flow, or the dipole
moment of the peculiar velocity field.  For an idealized
densely-sampled survey, the bulk flow would reflect the gravitational
pull of mass near to and beyond the survey limits.  It is therefore
sensitive to the distribution of mass on the largest scales.

We fit the radial components of the cluster peculiar velocities with
the flow model
\begin{equation}
U(\bvr{r}) = \bvr{V}\cdot\rhat + \frac{\Delta H}{H} \, \bvr{r}
\label{eq:flowsimp}
\end{equation}
where the free parameters are $\bvr{V}$, the bulk flow vector, and
$\Delta H/H$, a perturbation to the assumed Hubble constant.  The
distances quoted in SMAC-IV are in units of \kms\ and have already
been adjusted so that the best fit gives $\Delta H = 0$, but leaving
this parameter free allows for the correct propagation of errors into
the bulk flow $V$.  We use Galactic coordinates as the basis for our
Cartesian coordinates: $z$ is toward the North Galactic Pole
($b=90\degr$), $x$ is in the direction of the Galactic centre ($l =
0\degr$), and $y$ is in the direction of rotation ($l=90\degr$).

As noted above, the errors in the peculiar velocities are not
independent but rather are coupled through the velocity-dispersion
system-matching corrections described above and in Paper III.  We
construct a $56 \times 56$ covariance matrix which consists of the
variance in distance due to scatter in the FP plus the square of a
``thermal'' scatter $\sigma\sbr{th} = 250\,\kms$ , where the latter
term reflects the small-scale noise of individual clusters around the
mean bulk flow.  These errors are independent from cluster to cluster
so this part of the covariance matrix is diagonal. This is then added
to the system-matching covariance matrix to obtain the full covariance
matrix, $C$ .  We then minimize
\begin{equation}
\chi^2 = \sum_{i,j} [u_i - U(\bvr{r}_i)]\,[C^{-1}]_{ij}\,[u_j -
  U(\bvr{r}_j)]
\label{eq:chisquared}
\end{equation}
where the $C^{-1}$ is the inverse covariance matrix.  

To assess the depth of a sparse sample such as SMAC, we define an
error-weighted effective depth,
\begin{equation}
R_* = \frac{\sum_i r_i/\sigma_i^2}{\sum_i 1/\sigma_i^2}
\end{equation}
where $\sigma_i^2$ are the peculiar velocity errors, i.e., the
diagonal elements of $C$.  For an idealized, densely-sampled survey
filling a top-hat sphere of radius $R$
with uniform distance errors for all objects, one would obtain $R_* =
R/2$.  With our choice of error weighting, the SMAC sample has $R_*
\sim 6300\,\kms$, as expected for a survey extending to $\sim 12000
\kms$.

\subsection{Results}

For our preferred choice $\sigma\sbr{th} = 250\,\kms$, the SMAC bulk
flow is $687\pm203$ \kms, toward \lberr{260}{13}{0}{11}.  In Galactic
Cartesian coordinates, this is $V_x = -124\pm172\,\kms$, $V_y =
-676\pm190\kms$, $V_z = -5\pm131\,\kms$, and in Supergalactic Cartesian
this becomes $V\sbr{SGX} = -367\,\kms$, $V\sbr{SGY} = 59\,\kms$,
$V\sbr{SGZ} = -578\kms$, in the direction SGL $=171 \degr$, SGB $=-57
\degr$.  This flow is significantly different from zero at the 99.91\%
confidence level (CL).
In some papers (\citealt{LauPos94}, including \citealt{HudSmiLuc99}),
the authors allow for ``error-bias''. This is a bias which arises
because the bulk flow amplitude is the square root of the quadrature
sum of vector components and so is a positive definite quantity. Even
if the true bulk flow were zero, random errors in the components would
yield a positive amplitude bulk flow. For the same reasons, in the
more general case of a non-zero bulk flow, random errors bias the
amplitude high.  The ``error-bias corrected'' value of the SMAC bulk
flow is 620 \kms.

\begin{figure*}
\epsfxsize\textwidth \epsfbox{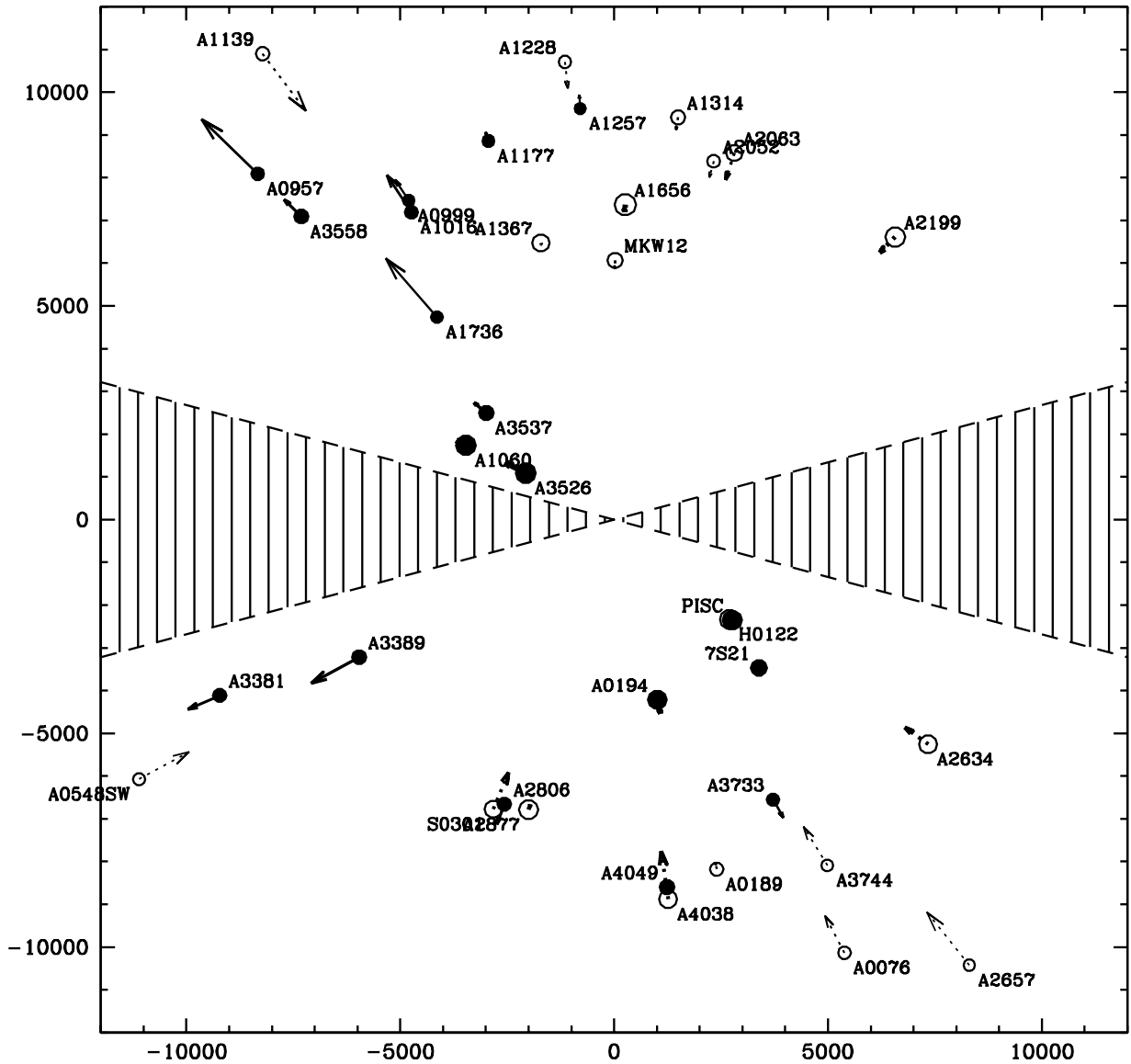}
\caption{ Peculiar velocities of SMAC clusters projected onto a plane
  in which the negative X-axis is along the direction of the SMAC bulk
  flow and the vertical axis points to the Galactic poles.  Clusters
  within $\pm45 \degr$ of the plane are plotted.  The circle indicates
  the estimated distance to the cluster, projected onto the plane, and
  the end of the tail is at the CMB-frame $cz$, so the peculiar
  velocity is indicated in \kms\ by the length of the tail. Clusters
  with peculiar velocities away from the origin are filled circles
  with solid tails, whereas clusters with peculiar velocities toward
  the origin are shown as open circles and dotted tails.  The size of
  the circle scales inversely with the distance errors.  The hatched
  region indicates the Zone of Avoidance ($|b| < 20\degr$).
\label{fig:smacflowplane}
}
\end{figure*}

\begin{figure*}
\epsfxsize\textwidth \epsfbox{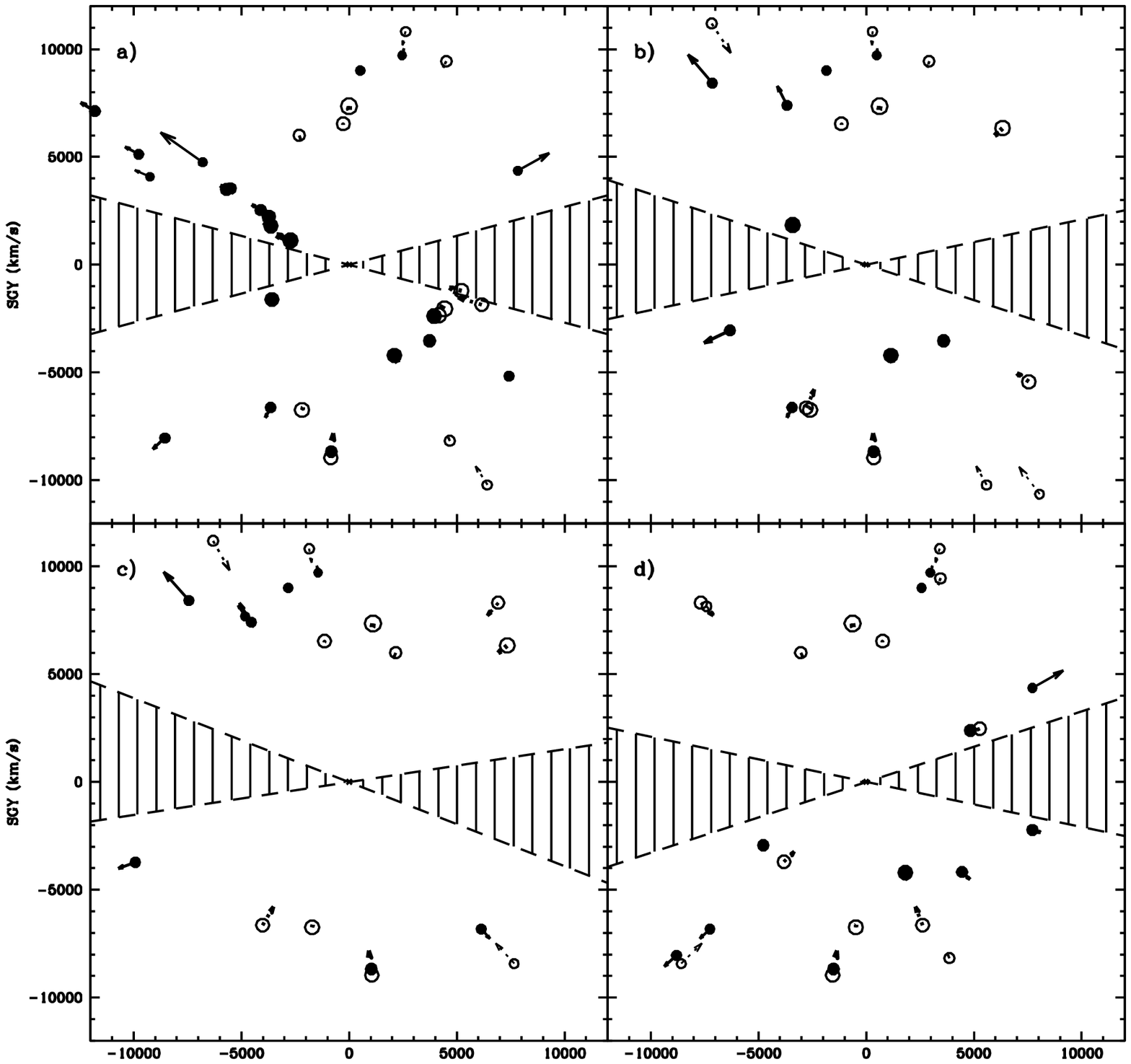}
\caption{ 
  Peculiar velocities of SMAC clusters with symbols as in
  \figref{smacflowplane}.  Four planes are shown: (a) the
  Supergalactic Plane; (b) a plane rotated by 45\degr\ from the
  Supergalactic plane around the the SGY axis. The horizontal axis is
  $1/\sqrt{2}$ SGX $+ 1/\sqrt{2}$ SGZ (toward SGL$=0$,SGB$=45\degr$).
  The SMAC flow is only 13\degr\ from this axis. The signature of a
  bulk flow, namely infalling objects on one half of the plot and
  outflowing ones on the other, is clearly seen in this panel. c) The
  SGX=0 plane d) a plane rotated a further 45\degr\ so that the
  horizontal axis is $1/\sqrt{2}$ SGX $- 1/\sqrt{2}$ SGZ.  Only
  clusters within $\pm22.5 \degr$ of the plane are plotted.
\label{fig:fourplanes}
}
\end{figure*}

\figref{smacflowplane} shows a ``tadpole diagram'' of the SMAC data
projected onto a plane in which the negative X-axis is along the
direction of the SMAC bulk flow and the vertical axis points to the
Galactic poles.  Note the excess of outflowing clusters on the
left-hand side and the inflowing clusters on the right-hand side, the
signature of a bulk flow.  In contrast, the objects at the top and
bottom scatter around zero peculiar velocity. The bulk flow has a
negligible component in the vertical (NGP/SGP) direction.
\figref{fourplanes} shows the SMAC data in four different planes
rotated by 45\degr around the SGY axis. Panel (a) corresponds to the
Supergalactic Plane.  Panel (b) is closest to the plane of the SMAC
flow.

\begin{figure*}
\epsfxsize\textwidth \epsfbox{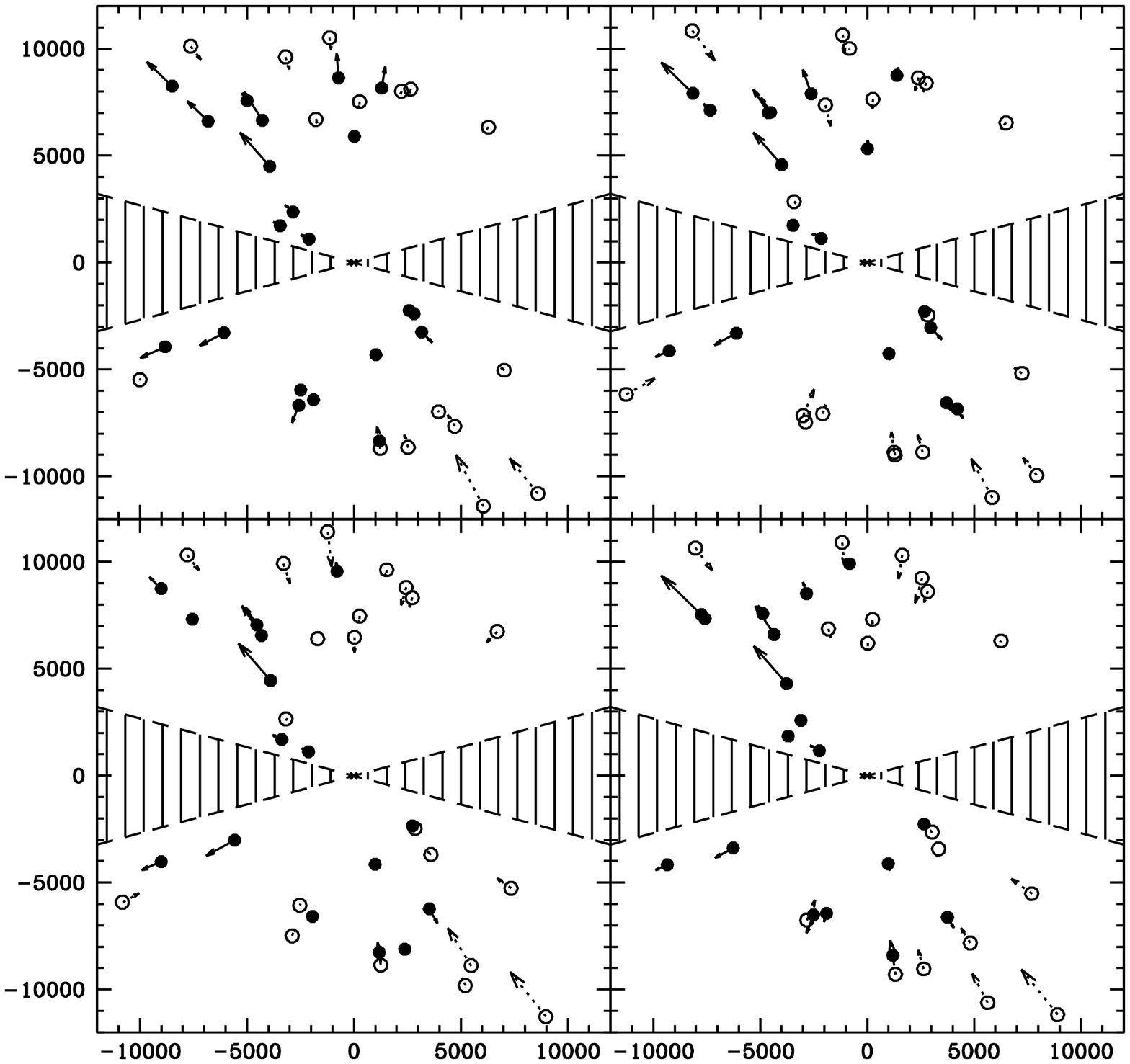}

\caption{
  As in \figref{smacflowplane}, but showing Monte Carlo realizations
  of the SMAC peculiar velocity field where we have perturbed the
  observed distances around their measured values according to their
  errors.  Notice that the outflow pattern on the left hand side and
  the inflow pattern in the lower right corner are robust to random
  errors.
\label{fig:monte}
}
\end{figure*}

It is difficult to illustrate the peculiar velocity errors in tadpole
diagrams. \figref{monte} shows four Monte Carlo realizations of the
peculiar velocity field where we perturb the observed distances around
their measured values with a Gaussian random error.  Robust regions of
the peculiar velocity field include the clump of positive peculiar
velocities in the top left quadrant and the negative peculiar
velocities in the distant part of the bottom right quadrant.

The bulk flow errors quoted above are marginal errors, i.e. they are
the diagonal elements of the covariance matrix.  The error covariance
matrix is not isotropic; it is a triaxial ellipsoid.  The long axis of
the error ellipsoid is oriented toward \lb{223}{-10} (and its
antipode).  The error along this direction is 222 \kms. The
intermediate axis is toward \lb{332}{-41}, and the error is 152
\kms. The error along the short axis is 102 \kms. The corresponding
direction (toward \lb{312}{46}) is the direction along which the bulk
flow is most precisely measured.  Because the SMAC sample has good sky
coverage, the bulk flow is almost independent of the monopole ($\Delta
H/H$) term: the correlation coefficient between it and the bulk flow
amplitude is only 0.09.

We find that the recovered SMAC bulk flow depends very weakly on the
value of $\sigma\sbr{th}$.  If we fit for $\sig\sbr{th}$ we find
$\sigma\sbr{th} = 190\pm105\,\kms$.  For $\sig\sbr{th} = 190\,\kms$, the
bulk flow is only 16 \kms\ lower than for the default $\sig\sbr{th} =
250\,\kms$ case.

The flow therefore appears to be quite cold, i.e. the bulk flow
dominates over small-scale ``thermal'' motions of clusters. The cosmic
Mach number \citep{OstSut90} is the ratio of the bulk flow velocity of
a volume to the 3D velocity dispersion of the objects in the frame of
the mean motion, ${\cal M} = |V|/(\sqrt{3}\,\sig\sbr{th})$.  For the
SMAC sample, we find ${\cal M} = 2.0\pm1.3$ The errors on this
quantity are obtained by propagation of errors, a procedure not
strictly valid given the size of the errors in comparison to the
measurement.

\begin{figure}
\epsfxsize\columnwidth \epsfbox{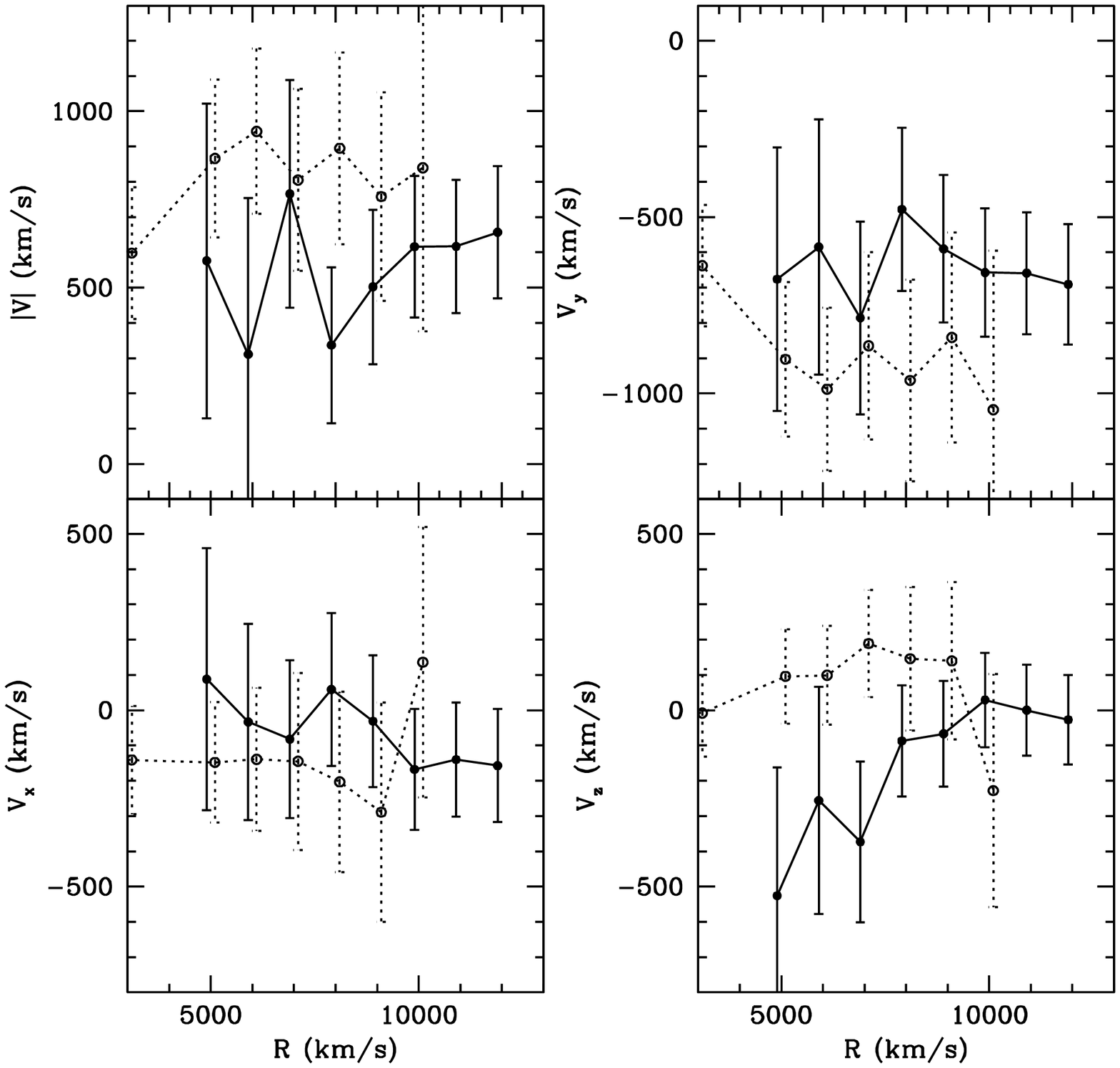}
\caption{ Bulk flow for different subsamples of the SMAC data selected
by distance.  Solid curves indicate subsamples with data in an inner
sphere extending from the LG out to $R$. Dotted curves show bulk flows
of subsamples in an outer shell extending from radius $R$ to the edge
of the SMAC sample.  For clarity, the curves are offset slightly in
the horizontal direction. For a given curve, the data points at
different $R$ are not independent, but at each $R$, the two subsamples
are statistically independent.  In all cases, the bulk flows of inner
and outer samples are consistent with each other.
\label{fig:innerouter}
}
\end{figure}

To test the coherence of the bulk flow, we have divided the sample
into statistically-independent nearby ($z_c < 7500\,\kms$, with an
effective depth $R_* = 50 \hmpc$) and a distant subsamples ($z_c >
7500\,\kms$, with $R_* = 98\,\hmpc$).  The nearby sample has a bulk
flow of $640\pm290\,\kms$ toward \lb{266}{-17}. This is significantly
different from 0 at the 97\% confidence level. At this depth, flows of
order several hundred \kms\ are known to exist
\citep{CouDek01,Zar02b}, and the SMAC result is consistent with
these. The distant subsample yields a somewhat larger bulk flow
($975\pm300\kms$ toward \lb{267}{16}).  Although the errors are large,
the bulk flow of the distant subsample is significantly different from
zero at the 99.5\% confidence level
It is not statistically different from the bulk flow of the
nearby subsample.  \figref{innerouter} shows the components of the
bulk flow for nearby and distant subsamples separated at different
values of $R$.  At no value of $R$ are the bulk flows of the two
subsamples inconsistent.  This coherence suggests that there are
significant contributions to peculiar velocities in the nearby
Universe arising from density fluctuations on very large ($\sim 100
\hmpc$) scales.

\subsection{Possible Sources of Systematic Error}
\label{sec:sys}

In this section, we examine possible systematics and assess the
robustness of the bulk flow. There are many sources of error that
could affect a cluster's peculiar velocity, but in general such errors
will only increase the scatter in the FP or add a random error to the
cluster peculiar velocities.  The bulk flow is the dipole of the
peculiar velocity field, so in order for it to be affected by a
systematic error, the systematic error must be coherent over large
areas of sky.  The most likely sources of such an effect are problems
matching data from different observing runs, or systematic errors in
Galactic extinction corrections.

\subsubsection{Spectroscopic Data}
\label{sec:specsys}

Systematic errors in velocity dispersion measurements could, in
principle, lead to large errors in the bulk flow.  For example, if
velocity dispersions measured in the North were systematically 1\%
smaller than those measured in the South, then distances in the North
would be too short by 1.4\%. For the SMAC sample, this would yield a
spurious bulk flow of $\sim 50\,\kms$ toward the North Celestial Pole.
Because of these possible effects, great care was taken to obtain
comparison data for the same objects from multiple telescopes
(SMAC-I).  In SMAC-III, we calculated the corrections required for
each run as well as the correlated uncertainties in these corrections.
The latter are fully propagated to the bulk flow and are thus included
in the error covariance matrix appearing in \Eqref{chisquared}.
Because of the large number of comparison data, these errors do not
dominate the error in the bulk flow: had we neglected these, we would
have quoted an error of 188 \kms (compare with the correct value of
203 \kms). Subtracting these in quadrature, we can quantify the error
in the bulk flow from system-matching uncertainties: 76 \kms.

\begin{figure}
  \epsfxsize\columnwidth\epsfbox{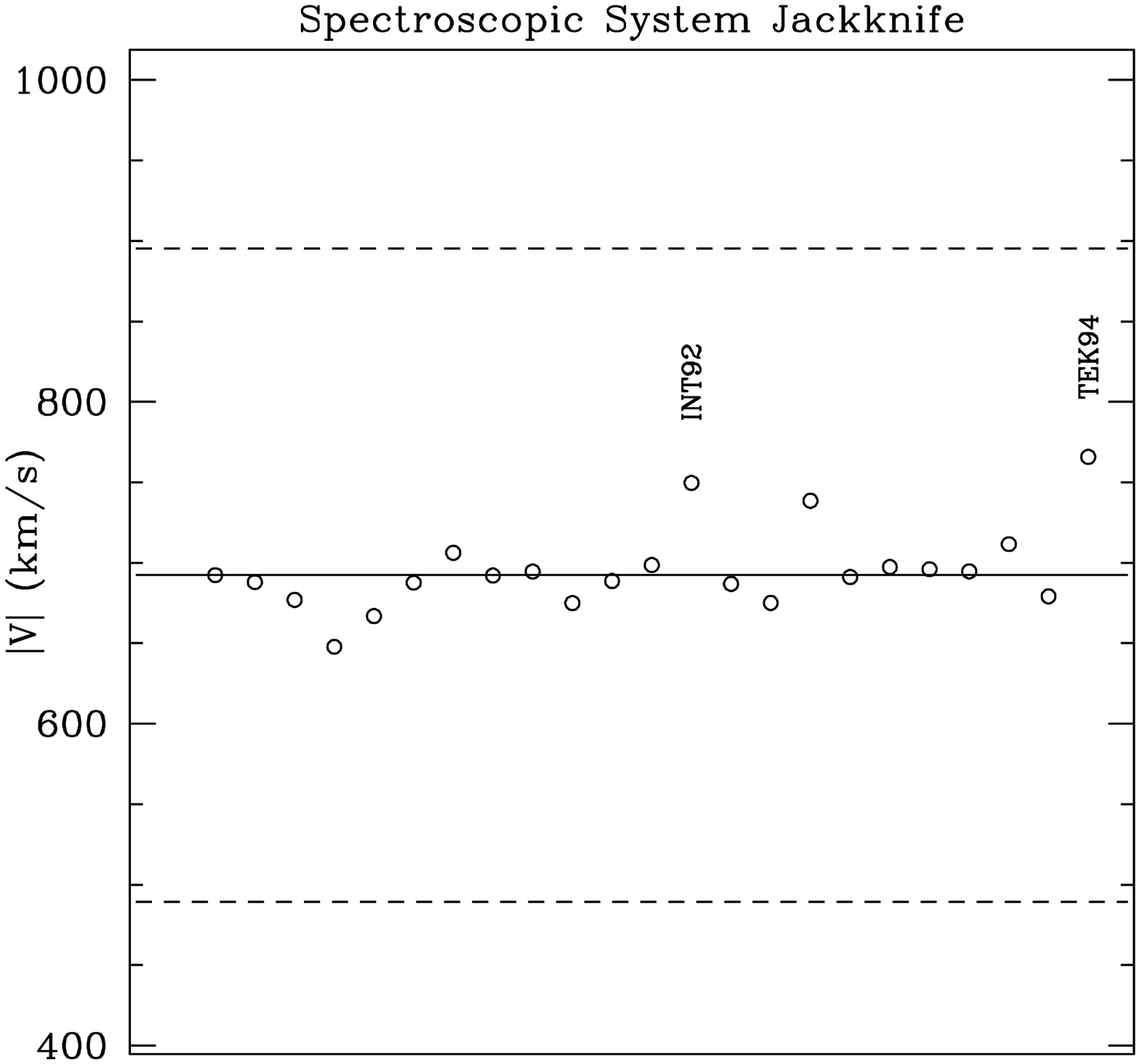}
\caption{ The jackknife test applied to the bulk flow statistic
  showing the effect of excluding a given spectroscopic run from the
  full solution. The vertical axis shows the amplitude of the bulk
  flow and the horizontal axis indexes different spectroscopic runs.
  Open circles show results for each spectroscopic run in turn.  Those
  runs whose exclusion changes the bulk flow by more than 50 \kms\ are
  labelled. For details of the labels, see \citet{SmiLucHud00}. The
  solid line shows the best fit result from the whole sample, and the
  dashed lines show the 1 $\sigma$ errors on the amplitude.  No
  individual spectroscopic run has a significant effect on the bulk
  flow.
\label{fig:spectrojknf}
}
\end{figure}

In order to examine in more detail the effects of individual
spectroscopic datasets, we have performed a ``jackknife'' test in
which we exclude each dataset in turn, and recalculate the bulk flow.
Some clusters, however, were observed only in one run, so the
jackknife removes entire clusters from the sample and consequently
changes the spatial sampling of the survey.  The results of this test
are shown in \figref{spectrojknf}.  No single spectroscopic dataset
has an effect on the bulk flow at a level of more than 70 \kms.

We conclude that errors in spectroscopic systems are fully quantified
and controlled in the SMAC sample.

\subsubsection{Extinction Corrections}

The proximity of the SMAC flow to the Galactic plane suggests that
errors in the extinction corrections could affect the bulk flow.  In
this subsection, we investigate the effects of extinction on the SMAC
sample. Note that errors in distance due to random errors in
extinction are included in the peculiar velocity errors tabulated in
SMAC-IV, by assuming that the SFD extinctions are accurate to 16\%.
This error is applied to the cluster as whole rather than to
individual galaxies within the cluster, but is assumed to be
independent from cluster to cluster.  Extinction errors would have the
greatest effect on clusters with the highest extinction. In the SMAC
sample, there are 7 clusters (A0400, A0539, A0426, J8, A2657, A3526,
A3733) with mean $E(B-V) > 0.1$ mag.\ and hence $A_R > 0.26$ mag.  We
find, however, that when these 7 clusters are excluded, the bulk flow
drops by only 35 \kms.  Thus the SMAC bulk flow is not substantially
affected by random extinction errors.

Of greater concern is the possibility of coherent errors in the SFD
maps.  \citet{Hud99} tested the SFD maps using data for early-type
galaxies, which have a tight intrinsic \bvmg\ relation, from
\citet{FabWegBur89}. He found no strong evidence for systematic
dipolar errors in the SFD maps, and set an upper limit of 16\% to such
systematic errors.  \citet{BlaLucBar01}, using the $(V-I)$--\mg\
relation also found no evidence of a systematic dipole error.

As a further test of systematic extinction errors, we have re-computed
peculiar velocities and bulk flows using the extinction maps of
\citet[hereafter BH]{BurHei82} in place of those of SFD.  It is worth
noting that the BH corrections yield some anomalously large peculiar
velocities. For example, A2634 has a peculiar velocity of $-1200\pm390
\kms$ with the BH corrections, whereas with the SFD corrections its
peculiar velocity is $-710\pm370\kms$. Similarly, the peculiar
velocity of A2657 is $-3200\pm1200\,\kms$ with the BH corrections, and
$-1600\pm1200\kms$ with the SFD corrections.  Using the BH
corrections, we obtain a bulk flow of 592 \kms, a reduction of 95
\kms\ compared with the result obtained with the SFD corrections. Thus
uncertainties in Galactic extinction appear to affect the bulk flow at
a level less than half the random error.

\subsubsection{Stellar Populations}

Variations in age and metallicity will introduce extra scatter in the
FP.  If there are systematic differences from cluster-to-cluster, then
it is possible that the cluster peculiar velocities could be
significantly in error.  \mg\ can serve as an indicator of age and/or
metallicity.  For example, the models of \citet{Wor94} indicate that,
for a typical elliptical, a $-0.15$ change in metallicity at solar
abundance yields a \mg\ change of $-0.03$ and an R-band change of
$-0.13$ mag.  Similarly if we add a 0.10 mass fraction in the form of
an intermediate-age (5 Gyr) population, \mg\ changes by $-0.0083$
mag.\ and R-band light changes by $-0.10$ mag.

In SMAC-IV, we compared residuals from the \mg\--\sig\ relation with
residuals from the inverse FP on a cluster-by-cluster basis and found
no strong evidence for a correlation.

We also showed, however, that the FP-\mg\ relation, which includes
\mg\ as a parameter in addition to the usual 3 FP parameters, does
reduce the scatter in log \sig\ in the inverse FP.  We prefer not to
use the FP-\mg\ relation for our default solution for two reasons:
first, \mg\ is not available for all of our galaxies, and second, the
scatter in {\em distance\/} for the FP-\mg\ relation is actually
larger than for the inverse FP, because the $\log \re$ coefficient in
the FP-\mg\ relation differs from that of the FP relation.
Nevertheless, if there are systematic variations from
cluster-to-cluster, these can be corrected by including \mg\ in the
distance indicator.  However, when we compare the same sample of
clusters, we find that including the \mg\ term increases the bulk flow
by 261 \kms\ to 831 \kms\ towards lb{254}{-7}, while also increasing
the error in the bulk flow to 230\kms.

\subsubsection{Morphological Mix}

The SMAC sample contains both E and S0 galaxies.  In \citet[hereafter
PP-II]{HudLucSmi97}, we compared the FP relations of E ($T\le-4$) to
S0 ($T \ge -3$) types.  There we found a small and marginally
significant ($2\sigma$) offset in the zero-point, in the sense that E
types are observed to have larger velocity dispersions than S0
galaxies at fixed \re\ and \sbe. \citet{BlaLucTon02} also noted
differences between E and S0 in their comparison of FP and Surface
Brightness Fluctuation (SBF) distances. In SMAC-IV, we repeated this
analysis, this time with a free coefficient of the RC3 $T$-type.
There we found a small and marginally significant correction to the FP
predicted velocity dispersions ($-0.0043\pm0.0019$ in $\log \sigma$
per unit of $T$-type).  This corresponds to E types having
$2.7\pm1.3\%$ larger velocity dispersions at a given \re\ and \sbe\
than S0 galaxies.  This difference is smaller than, but consistent
with, the offset found in PP-II.  We have not included this term in
our default solution because we do not expect any significant
difference in the ratio of Es to S0s across the sky.  When the term is
included, the bulk flow drops by only $\sim 30$ \kms.  If we treat E
and S0 subsamples separately, they both give a consistent bulk flow,
the E sample of 417 galaxies yields a slightly lower amplitude
($601\pm195\kms$) bulk flow than the S0 sample of 277 galaxies
($762\pm294\,\kms$). The larger bulk flow error for the S0-only sample
arises partly from the smaller number of galaxies in the S0 sample,
and partly from the larger scatter in the S0 FP relation.

\subsubsection{Outlying Data}

In SMAC-IV, we removed 5 galaxies which deviated by more than 3 sigma
from the best-fitting FP.  Re-including these 5 outliers reduces the
bulk flow by 56 \kms.

\begin{figure}
\epsfxsize\columnwidth \epsfbox{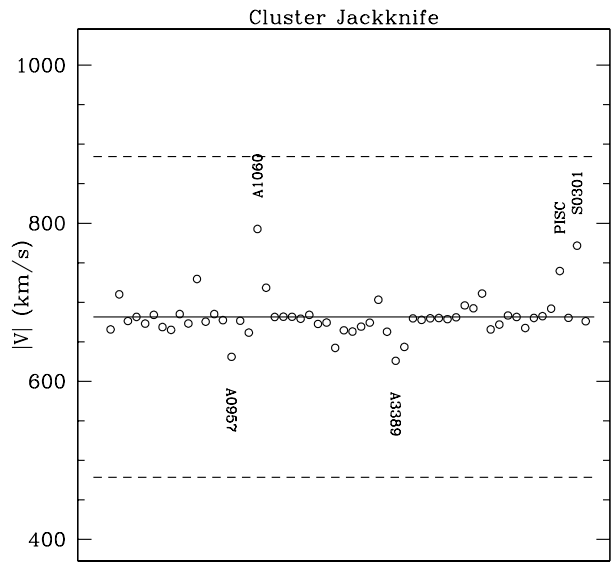}
\caption{ As in \figref{spectrojknf}, but showing the effect of
excluding individual clusters of galaxies.
\label{fig:clusjknf}
}
\end{figure}

In \figref{clusjknf}, we show the effect on the bulk flow amplitude of
removing entire clusters.  Clusters which affect the bulk flow more
than most are indicated.  No single cluster influences the bulk flow
upwards by more than 55 km/s. In particular, removing the nearby
Centaurus cluster (which includes both the Cen30 and Cen45
subcomponents) has little impact on the bulk flow (reducing it by only
40 \kms).

\citet{GibFruBot01} have suggested that FP clusters with large
internal FP scatter have significantly larger peculiar velocity
amplitudes. This might arise, for example, in cases where it is
difficult to separate double clusters in which the two components are
superimposed along the line of sight but are at slightly different
distances.  The spread of distances would then be increased, and the
mean distance, and hence peculiar velocity, would be more susceptible
to how the two components are sampled.  Classic cases where this
situation occurs are Abell 400 and Cen30/45.  We examined this issue
in Paper IV and found no strong evidence for a difference between the
peculiar velocities of high and low scatter clusters.

Despite our non-detection of this effect, we have experimented with a
weighting scheme in which the error for each cluster's distance is
based on the FP scatter for that cluster, rather than a global
value. This procedure down-weights those clusters with large internal
scatter.  When we apply this weighting scheme, we find that the bulk
flow drops, but only by $\sim 80\,\kms$.

\subsubsection{Malmquist Bias Corrections}

The procedure we have followed involves first calculating cluster
distances from the inverse FP and then fitting these to a flow model.
In the terminology of \citet{StrWil95}, this is ``Method I''.  In this
method, the estimate of each cluster's distance is its inverse FP
distances (as opposed to its redshift), and so these distances must be
corrected for Malmquist bias.  One advantage of working with clusters
is that this correction is small, since it scales as the inverse of
the number of objects in the cluster.

It is also possible to fit the flow model and FP relation
simultaneously. In this alternative ``Method II'' fit, redshift
(corrected by the flow model) is the {\em a priori\/} distance
indicator. As a result, no Malmquist bias are corrections necessary%
\footnote{Technically, there is a Malmquist-like bias associated with
the error in the estimated redshift, but at the distances considered
in this paper, this bias is very small.}.  A Method II analysis of the
SMAC sample assuming a simple bulk flow model yields a bulk flow
amplitude only $\sim 20\,\kms$ less than our standard result with
Malmquist bias corrections, indicating that the details of Malmquist
bias correction have little effect on the bulk flow.

\subsubsection{Summary of Systematic Effects}

We have examined several possible sources of systematic error. The
most obvious source of systematic error, namely mismatches between
velocity dispersions obtained on different runs is fully accounted for
in our standard error analysis. The only systematic effect which might
reduce the bulk flow is using the BH extinction maps (at the price of
introducing very large peculiar velocities for two clusters). This
would lower the flow by less than 100 \kms.  Thus systematic errors in
the bulk flow appear to be negligible in comparison to the random
error of $\sim 200\,\kms$.

\section{Other Flow Models}
\label{sec:othermod}

\subsection{Hubble Bubble}

\citet{ZehRieKir98}, in an analysis of 44 SNIa distances, claimed
evidence for a ``Hubble bubble'', i.e. that the local Universe within
7000 \kms\ is underdense with the result that the local value Hubble
constant is to high compared to the global value, $\Delta H/H =
6.5\pm2.2 \%$.  \citet{GioDalHay99}, using the TF relation, found a
statistically insignificant ``bubble'' of $\Delta H/H = 1\pm2 \%$.

We have tested this model with the SMAC data. We measure a local
Hubble bubble of $2.3\pm1.9$\% within the same distance studied by
Zehavi et al.  This result is consistent with zero. However, allowing
for errors in the Zehavi et al. result as well as ours, the two
results are consistent. We do not find any evidence of a Hubble bubble
if we cut the SMAC sample at other radii. Taking all of these results
at face value suggests there is at most a modest Hubble bubble with
$\Delta H/H \sim 3\pm1.3\%$.

\subsection{Bulk Flow plus Shear}

The large scale shear of the peculiar velocity field offers an
opportunity to identify the distances of the sources responsible for
the bulk flow \citep{LilYahJon86} particularly if they are near the
survey limits or in the Zone of Avoidance. For example, if there were
an attractor just beyond the SMAC survey limits, then, relative to a
pure bulk flow, the peculiar velocities would be more positive at the
edge close to the attractor and less negative on the opposite side of
the sky. Thus the residuals from the constant bulk flow would have a
tidal pattern, which might be measureable.  Since the strength of the
tidal field is inversely proportional to distance cubed, then if the
attractor was very far distant, the measured shear would be small.

To measure the shear, we modify \eqref{flowsimp} as follows:
\begin{equation}
U(\bvr{r}) = \bvr{V}\cdot\rhat + \rhat \cdot \bvr{h} \cdot \bvr{r}
\end{equation}
where $\bvr{h}$ is a symmetric shear tensor.  The shear tensor
measures anisotropies in the Hubble expansion.  When the shear tensor
is diagonalized, each eigenvalue corresponds to the difference in the
expansion rate ($\Delta H/H$) along the corresponding eigenaxis.  This
differs from \eqref{flowsimp} where the Hubble constant was allowed to
float, but was forced to be the same in all directions.

When this fit is performed, we find no significant reduction in
$\chi^2$ compared with bulk flow only fits.  The direction of
expansion ($6\pm3$\%) is towards \lb{315}{-7} and its antipole.  The
direction of expansion is very close to the negative SGY axis in the
Supergalactic Plane. The direction of compression ($5.5\pm3$\%) is
towards \lb{208}{-68} and antipole.  However, these amplitudes are not
statistically significant, and consequently the directions of
expansion and compression are ill-defined due to the large errors.
With this fit, the bulk flow is essentially unchanged: we find
$615\pm211$\kms\ toward \lb{259}{4}.

\subsection{Attractor Models}
\label{sec:toy}

Next we consider models with simple attractors.  Although such models
are likely to be only a crude approximation to a Gaussian random
density field, they allow us to gain some insight into cosmographical
features in the nearby Universe. The SMAC cluster data are rather
sparse so we do not attempt to identify attractors in the data
themselves. Instead we concentrate on attractor models using distances
and profiles fixed by other authors.

We model the radial infall component towards an attractor with the
following functional form:
\begin{equation}
u\sbr{a}(\bvr{r}) = \left(v\sbr{a} \frac{\bvr{r}\sbr{a}}{d\sbr{a}}
\left[\frac{(d^2\sbr{a}+c^2\sbr{a})}
{(r\sbr{a}^2+c\sbr{a}^2)}\right]^{(n\sbr{a}+1)/2} \right)\cdot \rhat \,,
\label{eq:attractorflow}
\end{equation}
where $d\sbr{a}$ is the distance from the LG to the attractor,
$\bvr{r}\sbr{a} = \bvr{d}\sbr{a}-\bvr{r}$ is the vector from the point
$\bvr{r}$ to the centre of the attractor, $c\sbr{a}$ is a core radius
and $v\sbr{a}$ is the velocity infall towards the attractor at the
position of the LG. The infall is $u\sbr{a} \sim 0$ at $r\sbr{a} \sim
0$, peaks near $r\sbr{a} \sim c\sbr{a}$ and then falls of as $u\sbr{a}
\sim r\sbr{a}^{-n\sbr{a}}$ at large $r\sbr{a}$.  Here we fix $n\sbr{a}
= 2$.

The first attractor we consider is the ``Great Attractor'', using the
flow model of Faber and Burstein (1988, hereafter FB88), except that
for simplicity we use $n\sbr{a} = 2$ rather than $n\sbr{a} = 1.7$
adopted by FB88. We keep the location ($r\sbr{GA} = 4200\,\kms$ in the
direction \lb{309}{18}) and core radius ($1400\,\kms$) of this
attractor fixed with the parameters in the FB88 model, and fit only
for one free parameter: the infall towards the GA at distance of the
LG.  With no bulk flow in the fit, the infall from the GA is $88\pm68
\,\kms$.  This is consistent with zero and differs significantly from
the result of FB88 who found $v\sbr{GA} = 535\,\kms$.  Note that in
\figref{fourplanes}a, there is no evidence of infall into the GA on its
the far side. We caution, however, that our data are sparse in the
immediate vicinity of the GA.  There are only three SMAC clusters for
which the FB88 GA model predicts significant negative peculiar
velocities ($u\sbr{GA} < -500\,\kms$).
For each of these three clusters, the SMAC peculiar velocity is
positive.  If we include a bulk flow in the fits, the GA infall is
even smaller ($13\pm70\,\kms$).  These conclusions do not change if we
adopt the GA position of \citet{KolDekLah95}.  We conclude that there
is little evidence from our sample for a very large mass at the
position of the GA. However, it should be noted that the errors are
large: the 95\% upper limit from our bulk flow plus GA model is
$v\sbr{GA} \sim 150\,\kms$.

\citet{Hud93}, in his analysis of the density field of
optically-selected galaxies, found the GA overdensity was better
described as a broad overdensity centred on the Centaurus cluster. He
predicted an infall of $287\pm62 \kms$ at the LG for $\beta\sbr{opt} =
0.5$.  In their analysis of SBF data, \citet{TonBlaAjh00} fitted a
``GA'' model, the centre of which is almost coincident with Centaurus
cluster. They obtained an infall of $289\pm137\,\kms$ at the LG, in
good agreement with the \citet{Hud93} prediction. If we adopt the
GA/Centaurus attractor centre of \citet{TonBlaAjh00}, we find an
infall of $58\pm153 \kms$.  It is difficult to make an exact
comparison because the details of the infall model of
\citet{TonBlaAjh00} is different from ours, but our results do not
appear to be in conflict with theirs.

It is clear that local ($R < 60 \hmpc$) attractors are not responsible
for the large-scale flow seen in the SMAC sample. The sample must be
responding to sources at distances beyond the SMAC effective depth of
6000 \kms. The next attractor considered is the Shapley Concentration
(SC) centered on the rich cluster Abell 3558 at \lb{312}{31} and a
distance of 145 \hmpc.  Since there are SMAC clusters quite close to
the SC, it is important to model the SC as an extended mass
distribution.  The model given by \eqref{attractorflow} allows for a
core radius of the attractor.  The SC core radius is not well
determined by the SMAC data themselves (a fit yields $\sim 30\pm20
\hmpc$) so we fix it at 30 \hmpc.

If we fit for only a SC attractor with no bulk flow, we find an infall
at the LG of $200\pm60\kms$. This corresponds to an excess mass of
$9\pm3\ten{16}(\om/0.3)^{0.4} h^{-1} M_{\sun}$ from the SC region.

If we fit for the SC attractor plus a bulk flow, the infall due to the
SC reduces to $140\pm80\kms$, and so is not statistically
significant. For this model, the bulk flow is $620\pm220\kms$ and the
direction has swung around by $\sim 10\deg$ to \lb{248}{-6}. This bulk
flow is inconsistent with zero at the 95\% CL.  Note that there is
strong covariance between the SC infall and the bulk flow which boosts
both of their marginalized errors.

We conclude that there is a tantalizing suggestion for substantial
mass at the SC, but that it is not the sole source of the SMAC bulk
velocity.

\section{The peculiar velocity field predicted by the \emph{IRAS} PSCz
  survey}
\label{sec:pscz}

\subsection{Method}

The toy attractor models discussed above are rather crude. A better
approach to modeling the peculiar velocity field is to use a redshift
survey to reconstruct the real space density field of galaxies.  If
mass density contrasts are related to galaxy number-density contrasts
according to a simple biasing relation, e.g. $\delta\sbr{g} = b
\delta$, then we can obtain predictions for the peculiar velocity
field using the linear theory equation
\begin{equation}
\bvr{u}\sbr{lin}(\bvr{r}) = \frac{\beta}{4\pi} \int
\delta\sbr{g}(\bvr{r}')\frac{\bvr{r}'-\bvr{r}}{|\bvr{r}'-\bvr{r}|^3}
d^3\bvr{r}'\,,
\label{eq:pecvellin}
\end{equation}
where $\beta = \omsix/b$. This method has been used by a number
of workers to compare observed and predicted peculiar velocities and
hence obtain $\beta$ (see reviews by \citet{StrWil95},
\citet{CouDek01} and \citet{Zar02b}).  For predictions of the peculiar
velocities of clusters of galaxies, linear theory is adequate.

While the integral in \eqref{pecvellin} extends over all space, in
practice, redshift surveys typically do not have data in the Zone of
Avoidance and are truncated at large distances where the corrections
for selection effects become large.  In the case of a distance-limited
redshift survey, \eqref{pecvellin} then yields peculiar velocities in
the frame of the center of mass of the redshift survey volume. In
general, there will be contributions to the CMB-frame peculiar
velocity from sources outside the redshift survey volume.  If the
redshift survey is much deeper than the peculiar velocity survey, and
there are no massive structures in the ZOA, then the tidal or
quadrupole effect from distant attractors, which falls as $r^{-3}$,
will be small.  It is then sufficient to model these residual
contributions by adding a simple dipole term $\vext$ to
\eqref{pecvellin}:
\begin{equation}
\bvr{u}(\bvr{r}) = \bvr{u}\sbr{lin}(\bvr{r}) +
\vext\cdot\hat{r} \,.
\label{eq:psczbulk}
\end{equation}

There are several approaches to fitting $\beta$.  One common approach
is to perform the comparison between predicted and observed peculiar
velocities in the LG frame. In this comparison, the exterior dipole
cancels from the predictions since $\bvr{u}\sbr{LG}(\bvr{r}) =
\bvr{u}(\bvr{r}) - \bvr{u}(0)$. An alternative approach is to fit in
the CMB frame, but this ignores the LG as a data point entirely.

In this paper, we make the comparison in the CMB frame but we force
our solutions to be consistent with the LG's peculiar velocity. We
implement this by including the three Cartesian components of the LG's
peculiar velocity as if they were three additional peculiar velocity
data with zero observational error.  We do, however, allow both the LG
peculiar velocity components and the peculiar velocities of the SMAC
clusters to have a random ``thermal'' component to their error. In
this context, the ''thermal'' component represents the combined
effects of non-linearities in the peculiar velocity field and errors
in the peculiar velocities predicted from the galaxy density field.

\subsection{Application to PSCz}

For the galaxy redshift survey, we have used the \emph{IRAS} Point
Source Catalogue Redshift (PSCz) survey \citep{SauSutMad00} with
peculiar velocity predictions from \citet{BraTeoFre99}.  The distance
limit of the density field is 20000 km/s. The Galactic Plane ($|b| < 8
\degr$) is filled by interpolating the data at higher $|b|$.  The
``real space'' distances are obtained in a self-consistent way by an
iterative method (see \citet{BraTeoFre99} for additional details).

The flow model uses \eqref{psczbulk} where $\bvr{u}\sbr{lin}$ are the
\emph{IRAS} predictions, obtained by using the \emph{IRAS} galaxy
density field as $\delta\sbr{g}$ in \eqref{pecvellin}.  This model has
four free parameters: $\beta\sbr{I}$ and the three components of the
$\bvr{V}\sbr{ext}$.  We set the $\sigma\sbr{th} = 150\,\kms$ for this
comparison.

One final correction is necessary. We are making predictions for
clusters of galaxies, which are significant mass concentrations.  In
the \emph{IRAS} density field, these clusters will be located at a
given ``real-space'' position as obtained by the self-consistent
iterative reconstruction method. Due to distance errors, the observed
SMAC cluster will, in general, be located a different distance along
the line of sight from the \emph{IRAS} cluster.  We do not want the
\emph{IRAS} cluster to generate a spurious peculiar velocity at the
location of the SMAC cluster, so for each SMAC cluster we identify and
temporarily delete its counterpart in the \emph{IRAS} density field
before calculating the predicted peculiar velocity.

\subsection{Results}

\begin{table*}
\caption{Fits using the PSCz predicted peculiar velocity field 
\label{tab:pscz}}
\begin{tabular}{lrrrrrrr}
\hline
Fit & $\beta\sbr{I}$ & $|\vext|$ & $l$ & $b$ & P($\vext=0$) & $\chi^2$ & DOF
\\ \hline 
{\bfseries With LG} 
& $\mathbf{0.39\pm0.17}$ 
& $\mathbf{372\pm127}$ 
& \bfseries 273 
& \bfseries 6 &  \bfseries 0.02 & \bfseries 57.5 &  \bfseries  54  \\
With LG & $\equiv 0$ & $\equiv 0$ & & & & 95.1 & 58 \\
With LG & $\equiv 0$ &  $541\pm104$  & 269 & 11 & $4\ten{-7}$ & 62.7 & 55  \\ 
With LG & $0.70\pm0.13$ & $\equiv 0$ & & & & 67.3 & 57 \\
With LG & $\equiv 0.5$ & $327\pm104$ & 276 & 4 & $8\ten{-3}$ & 57.9 & 55 \\
No LG & $0.87\pm0.31$ & $637\pm184$ & 263 & -12 & $1\ten{-3}$ & 53.0 & 51 \\ 
No LG & $\equiv 0$ & $\equiv 0$ & & & & 80.4 & 55 \\
No LG & $\equiv 0$ & $661\pm182$ & 261 & -2 & $2\ten{-4}$ & 61.0 & 52 \\
No LG & $0.99\pm0.28$ & $\equiv 0$ & & & & 68.7 & 54\\
\hline
\end{tabular}
\end{table*}

\begin{figure}
\epsfxsize\columnwidth \epsfbox{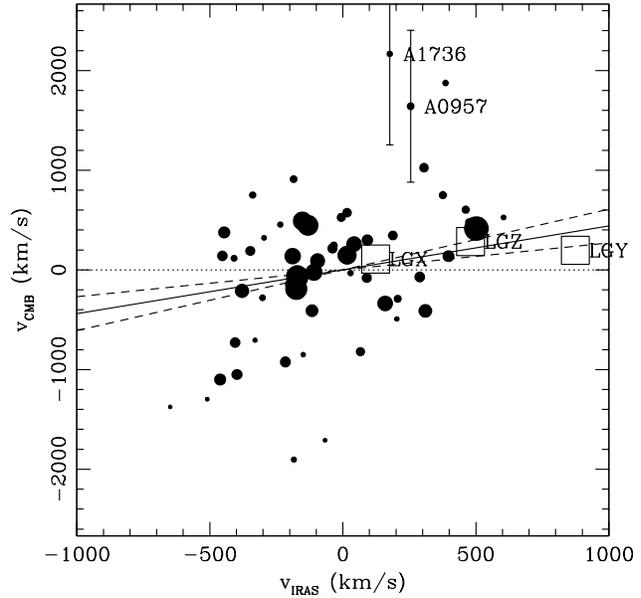}
\caption{ A comparison of observed peculiar velocities and predictions
  from the PSCz density field with $\beta\sbr{I} = 1$.  SMAC clusters
  are indicated by filled circles. Clusters which deviate by more than
  $2\sigma$ from the model predictions have error bars plotted and are
  labelled.  Vectorial components of the LG's peculiar velocity are
  indicated by open squares and are labelled. The best fitting
  external bulk flow $\vext$ has been subtracted from the
  observations.  Symbol sizes scales inversely with observational
  error. The solid line shows the $\beta\sbr{I}$ of best fit; dashed
  lines indicate the errors.
\label{fig:pscz}
}
\end{figure}

Simultaneous fits of the parameters in \eqref{psczbulk} are given in
\tabref{pscz}. The best fit has $\beta\sbr{I} = 0.39\pm0.17$ and
$\vext = 372\pm127\,\kms$ towards \lberr{273}{17}{6}{15}.
\figref{pscz} shows $u\sbr{lin}$ (with $\beta\sbr{I} = 1$) vs.\
$u\sbr{obs}$. The slope of the line of best fit then gives
$\beta\sbr{I}$.

\begin{figure}
\epsfxsize\columnwidth \epsfbox{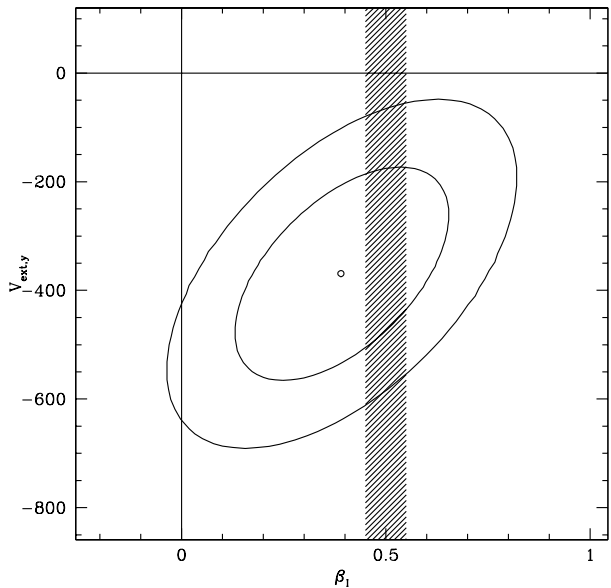}
\caption{ Parameter fits to the SMAC+LG peculiar velocity data using
the PSCz predictions. Contours show the 68\% and 95\% confidence
contours on $\beta\sbr{I}$ and the $y$-component of $\vext$. The open
circle shows the best fit values.  The hashed vertical bar indicates
the concordance value of $\beta\sbr{I}$ discussed in \secref{extbeta}.
\label{fig:pscz_ellipse}
}
\end{figure}

For our best fit, $\chi^2 = 57.5$ for 54 degrees of freedom.  In
comparison, the null case of $\beta\sbr{I} = 0, \bvr{V}\sbr{ext} = 0$
has $\chi^2 = 95.1$ for 58 degrees of freedom.  The marginalized
errors quoted above are large because there is strong correlation
between the $\beta\sbr{I}$ and $\vext$ terms: as $\beta\sbr{I}$
increases, $|\vext|$ drops. The error contours associated with this
fit are shown in \figref{pscz_ellipse}.  The covariance complicates
the interpretation of these results. Of the two components of the flow
model, $\beta\sbr{I}$ is the more significant. If $\vext$ is set to
zero, $\beta\sbr{I} = 0.70\pm0.13$, a $>5\sigma$ detection of $\beta$.
This ``high-$\beta$'' model is disfavoured at the 98\% CL compared to
the model with $\beta\sbr{I}$ and $\vext$ free, for which $\chi^2$
falls by 10 with the removal of 3 degrees of freedom.  Finally, if
$\beta\sbr{I}$ is fixed at 0.5 (as external comparisons suggest, see
below), then we find $\vext = 327\pm104\,\kms$.

It is well known that \emph{IRAS} misses early-type galaxies in the
cores of clusters.  This effect may be particularly dramatic in
regions such as the Shapley Concentration. We have fit a flow model
with an attractor at the position of Shapley (core radius 3000 \kms)
in addition to the \emph{IRAS} predictions and an external bulk flow
$\vext$.  For this fit, we find an infall of $129\pm84$ \kms\ at the
LG. This is only $\sim 10\,\kms$ less than the infall into SC found
above (without the PSCz flow model), indicating that SC is very weak
in the PSCz.  For this fit the external bulk flow reduces to $300\pm
140\,\kms$.

\subsection{Discussion}
\label{sec:irasdiscuss}

The degeneracy between $\beta$ and $\vext$ make it difficult to
build a unique flow model from SMAC+LG data alone. There are several
independent methods of determining $\beta\sbr{I}$ and $\vext$.

\subsubsection{External constraints on $\beta\sbr{I}$}
\label{sec:extbeta}

\begin{enumerate} 

\item {\bf Matching the dipole to the LG's motion}.  This method
  consists of using \eqref{pecvellin} to predict $\bvr{v}(0)$ and
  adjusting $\beta\sbr{I}$ to obtain the best fit with the observed
  627 \kms\ motion of the LG. If the flow model allows for $\vext$ as
  well, then the two components are degenerate (one is fitting 3 data
  with 4 free parameters).  As a result for this method to yield
  constraints on $\beta\sbr{I}$ one has to make some assumptions for
  $\vext$.

  One approach is to allow cosmological models to predict the
  coherence between a distance-limited gravity dipole and the peculiar
  velocity of the LG. The most recent estimate of $\beta\sbr{I}$ by
  this method \citep{CieCho04} yields $\beta\sbr{I} =
  0.64^{+0.24}_{-0.11}$.
  
\item {\bf Fitting peculiar velocity data}. \citet{WilStr98} found
  $\beta\sbr{I} = 0.50\pm0.04$ by comparing the predictions of the
  \emph{IRAS} 1.2 Jy survey to the Mark III data \citep{WilCouFab97}.
  \citet{BraTeoFre99} found $\beta\sbr{I} = 0.42\pm0.04$ from the
  comparison of the PSCz to the SFI catalogue \citet{GioHayFre98}.
  \citet{BlaDavTon99} obtained $\beta\sbr{I}=0.42^{+0.10}_{-0.06}$
  from a comparison of Surface Brightness Fluctuation survey peculiar
  velocities to \emph{IRAS} 1.2 Jy predictions. These three analyses
  are conducted in the LG frame and are insensitive to \vext.
  \citet{ZarBraHof02} found $\beta\sbr{I} = 0.51\pm0.06$ from
  comparison of ENEAR \citep{daCBerAlo00} and SFI peculiar velocities
  to PSCz in the CMB frame.
  
\item {\bf Redshift-space distortions.} \citet{HamTegPad00} found
  $\beta\sbr{I} = 0.41_{-0.12}^{+0.13}$ from an analysis of
  redshift-space distortions in the PSCz survey. \citet{TayBalHea01}
  found essentially identical results.
  
\item {\bf Expectations from cosmological models.}  Using the WMAP,
  2dF and Lyman-$\alpha$ forest data, \citet{SpeVerPei03} found
  $\omsix\sigma_8 = 0.38^{+0.04}_{-0.05}$. Combining this with
  the measurement $\sigma\sbr{8,I} = 0.80\pm0.05$ from
  \citet{HamTeg02}, we obtain $\beta\sbr{I} = 0.48\pm0.06$.

\end{enumerate}

The ``concordance'' value $\beta\sbr{I}=0.5$ is consistent with all of
the above determinations at the $2\sigma$ level.
Note that the value $\beta\sbr{I} = 0.5$ requires a residual motion of
$\sim 160\,\kms$ towards \lb{319}{13} (or $V_x = 119$, $V_y = -110$,
$V_z = 35$) from outside the PSCz volume (i.e. from beyond 200 \hmpc\
or in the Zone of Avoidance) in order to match the observed peculiar
velocity of the LG.

\subsubsection{External constraints on $\vext$}

What is the expected value of $\vext$? There are four sources of
uncertainty in the PSCz predicted peculiar velocities.

\begin{enumerate}
  
\item {\bf Shot noise within $R \sim 200 \hmpc$}. At $R=150 \hmpc$,
  \citet{SchBraTeo99} quote errors of $160 \beta\sbr{I}\,\kms$ on the
  predicted velocity of the LG.  They do not extend their analysis to
  greater distances, but an eyeball extrapolation of their Figure 5
  suggests that this grows to $\sim 200 \beta\sbr{I}$ at $R=200
  \hmpc$.

\item {\bf Sources in the Galactic Plane}. \citet{SauDMeVal00}
  published preliminary results from the ``Behind the Plane'' (BTP)
  survey (an extension of the PSCz to lower Galactic latitudes).  The
  BTP dipole in units of \kms (assuming $\beta\sbr{I} = 0.5$), grows
  from $V_x = -60, V_y = -325, V_z = 350$ at $R \sim 100 \hmpc$, to
  $V_x = -60, V_y = -475, V_z = 350$ at $R \sim 200 \hmpc$.  The
  misalignment with the CMB dipole of only $\sim 13 \degr$. At $R \sim
  300 \hmpc$, the dipole is $V_x = -25, V_y = -675, V_z = 350$, or
  $\sim 761\,\kms$ towards \lb{268}{27}, and the misalignment is
  reduced to $\sim 8 \degr$.  The BTP dipole is in better directional
  agreement with the LG's observed motion than the PSCz dipole. It is
  possible that this alignment is fortuitous since the shot-noise
  effects (and systematics such as the ``rocket-effect''
  \citep{Kai87}) are likely to be large beyond 200 \hmpc.  There is a
  particularly strong change in $V_y$ in the range 180-240 \hmpc\
  which is present in both the PSCz and BTP dipoles, but is much
  stronger in the latter.  This suggest that there are dynamically
  important structures close to the Galactic Plane.

  For purposes of assessing $\vext$ result from the SMAC sample, an
  estimate of the effect of hidden sources in the Plane is given by
  the difference between the growth in the BTP dipole and the PSCz
  dipole at distances beyond the SMAC sample ($\sim 100 \hmpc$).  The
  difference in \kms\ for $\beta\sbr{I} = 0.5$ is $\Delta V_x = +105,
  \Delta V_y = -135, \Delta V_z = -20$, or 172 \kms towards
  \lb{308}{-7}.  We estimatethe error on this to be 50\% or 85 \kms.
  
  The ``Clusters Behind the Zone of Avoidance'' (CIZA), an X-ray
  selected survey galaxy clusters, \citet{EbeMulTul02} also indicates
  several massive clusters at low galactic latitudes.  However, these
  appear to be located either in the vicinity of the GA at 60 \hmpc\
  (A3627 and CIZAJ1324.7-5736) or nearer to SC at a distance of $\sim
  150 \hmpc$ (CIZA J1653.0-5943 and Triangulum Australis) and not at
  180-240 \hmpc\ suggested by the BTP dipole.  The gravity dipole of
  the CIZA sample \citep{KocEbeMul03} has a strong contribution at
  $150 \hmpc$ which is not seen in the PSCz or the BTP.

\item {\bf Sources beyond 200 \hmpc}. Since the PSCz/BTP data are
  noisy beyond 200 \hmpc, it is difficult to assess this empirically.
  For a $\Lambda$CDM cosmological model with $\Gamma = 0.21, \om=0.3,
  n=1$ and $\sigma_8 = 1$, we expect the rms contribution to the bulk
  motion in the nearby Universe arising from sources beyond 200 \hmpc\
  to be $\sim 50\,\kms$ in each component.
   
\item {\bf Missing contributions from high-density superclusters}.
  \emph{IRAS} does not detect early-type galaxies. For individual
  clusters this is a weak effect, which can be compensated by
  increasing $\beta\sbr{I}$. There are certain extreme regions, such
  as the SC and the Horologium-Reticulum superclusters, where there is
  an astounding excess of clusters.  For example, \citet{TulScaVet92}
  find 29 Abell/ACO clusters within 50 \hmpc\ of the centre of the SC.
  This corresponds to a mean overdensity $\delta_c \sim 5$ in
  clusters.  If $b\sbr{c}/b\sbr{I} \sim 4$, \citep{BraTeoFre99}, then
  we expect the PSCz overdensity to be $\delta_I \sim 1.25$, yielding
  an infall at the LG of $250 \beta\sbr{I}\,\kms$.  In fact, the
  observed PSCz overdensity is only 0.2, yielding an infall of only
  $40 \beta\sbr{I}\,\kms$.  A similar situation occurs for the
  Horologium-Reticulum supercluster. Tully et al.\ finds a cluster
  overdensity of $\sim 4$ on scales of 50 \hmpc\ but the PSCz
  overdensity is only 0.14.  Clearly linear biasing does not operate
  in these very high-density regimes.
   
  Because the cluster overdensity is likely an overestimate of the
  true mass density (even when scaled by $b\sbr{I}/b\sbr{c} \sim
  0.25$), and the PSCz overdensity an underestimate, one might expect
  the true situation to lie in between. For $\beta\sbr{I} = 0.5$, we
  will estimate a residual dipole of $\sim 50\pm25\,\kms$ in the
  direction of the SC, and, because it is at greater distance, half of
  that for Horologium-Reticulum.

\end{enumerate} 

Adding the BTP residual dipole (172 \kms\ toward \lb{308}{b=-7}), plus
contributions directed to SC (50 \kms) and HR (25 \kms), we obtain an
estimate of $\vext = 225\,\kms$ towards \lb{305}{-4}.  Most of this is
due to the extra Galactic Plane sources in the BTP dipole. Errors on
$\vext$ estimated to be 90 \kms (systematic uncertainty) and 75 (shot
noise) and 50 (residual beyond 200 \hmpc). Summed in quadrature, this
yields an error of 150 \kms.  Clearly, given the uncertainties, there
is no firm external evidence that $\vext$ is required. The fact that
the best estimate is an agreement with the measured direction and
amplitude of $\vext$ from the SMAC survey ($372\pm127\,\kms$ towards
\lberr{273}{17}{6}{15}) suggests that these external sources do exist.

\section{Comparison with Other Results}
\label{sec:comparison}

The bulk flow calculated in this paper is a statistic derived from the
SMAC sample.  This is not to suggest that the true flow field within
12000 \kms\ is a pure dipole bulk flow.  Indeed, a pure dipole
velocity field is unlikely given that the mass density field is a
Gaussian random field with fluctuation power on a range of scales.
Due to sparse sampling, the bulk flow of any peculiar velocity sample
will have contributions from the true bulk flow of the volume, as well
as from higher order multipoles. This effect was first emphasized by
\citet{WatFel95}. It would, therefore, be naive to compare the bulk
flow from different sparsely-sampled surveys without allowing for the
error introduced by sparse sampling.  Nevertheless, that has been the
approach taken in the past, and it has led to an exaggeration of the
differences between various studies.

A full statistical comparison between surveys, based on the three
components of the bulk flow and the full error covariance matrices
allowing for the above-mentioned sparse sampling effects, is beyond
the scope of the present paper. A preliminary comparison of the bulk
flows of SMAC, \citet{Wil99b}, LP, \citet{DalGioHay99b},
\citet[hereafter EFAR]{ColSagBur01}, and the SNIa compilation of
\citet{TonSchBar03} has been performed by \citet{Hud03}, who concluded
that there was no inconsistency between any of these surveys, except
possibly for LP which is inconsistent at the 94\% level.

Instead of the full vectorial comparison, in this paper we opt for a
simpler, illustrative comparison which considers simply the component
of the bulk flow projected along a given axis directed towards
\lb{300}{10}.  This ``concordance'' direction has been chosen because
it is close to the mean flow of the above-mentioned sparse surveys
\citep{Hud03}, is within a few degrees of the Supergalactic Plane, and
is also very close to the bulk motion originally reported for the 7S
sample by \citet{DreFabBur87}. In his analysis of the Mark II peculiar
velocity catalogue, \citet{Hud94b} found that the gravity of the local
density field to 8000 \kms\ could not account for a residual bulk
motion of $400\pm45\,\kms$ toward \lb{292}{7}.  If this model is
correct, then deeper surveys such as SMAC should have bulk flows close
to this residual motion.

For the SMAC sample, with the bulk flow fixed in the concordance
direction \lb{300}{10}, the best fitting amplitude is $400\pm120
\kms$.  The following sections compare our results with those from
from other surveys, focussing first on those based on sparse samples
on large scales .

\subsection{Large-scale Surveys}
\subsubsection{Brightest Cluster Galaxies} 

\citet{LauPos94} found a bulk flow of $689 \pm 178\,\kms$ towards
\lb{343}{+52} from 119 clusters using the a distance indicator based
on the photometry of brightest cluster galaxies (BCGs). The direction
of the SMAC bulk flow is nearly 90\degr\ away from that of LP, so
these two results appear to be in poor agreement.  Nor does LP agree
with results from other peculiar velocity surveys
\citep[EFAR]{DalGioHay99b}. The LP sample is denser than other all-sky
surveys on simuilar scales and so it is less affected by sampling
errors than other sparser surveys, including SMAC.  Therefore sparse
sampling is unlikely to be a source of disagreement between LP and
other surveys.

In SMAC-IV, we compared our cluster distances with those of LP. We
found that of 41 clusters in common, four were discrepant at greater
than the $2\sigma$ level: A262, A1060 (Hydra), A3381 and A3733. In all
cases, comparison with the observed redshifts indicates that the LP
distances are too long corresponding to BCG magnitudes whch are too
faint.

One possible explanation for this discrepancy is obscuration by dust
in BCGs.  \citet{LaivanLau03} examined dust in 75 of LP's 119 BCGs.
They found signs of dust in $\sim 38\%$ of BCGs. In some cases the
dust appears to be quite extended, and this could affect the LP
magnitudes.

There is evidence that galaxies classified by Laine et al.\ as having
filamentary or patchy dust are systematically too faint.  If we
compare the ratios of the estimated distance from the LP BCG method to
the distance estimated by assuming that the BCG is at rest \wrt\ the
CMB, we find that for 49 galaxies with no dust the distance ratio is
$0.98\pm0.02$.  Galaxies with only nuclear ($\la 1 \arcsec$) dust
disks (Laine code ``D''), dust spirals (code ``S''), or rings (code
``R'') have a distance ratio $0.92\pm0.04$.  These small scale
features are unlikely to affect the 10 \hkpc\ aperture magnitude which
corresponds to a much larger angular scale, typically tens of
arcseconds \citep{PosLau95}.  However, the distance ratios are higher
for galaxies with filamentary (``F'') dust ($1.06\pm0.06$), patchy
(``P'') dust ($1.12\pm0.08$) or both patchy and filamentary dust
($1.28\pm0.10$).  The latter category consists of only 4 galaxies, the
BCGs of A0262, A1060, A3698 and A3733.  Three of these (A0262, A1060
and A3733) clearly have very extended dust in gray scale images (fig.\ 
1 of Laine et al.).  These same three galaxies are extreme outliers in
the SMAC vs.\ LP comparison discussed above.

If we remove from LP's sample the 19 galaxies with dust classified as
filamentary or patchy (or both) in Laine et al.'s Table 2, we obtain a
bulk flow of $707\pm261\,\kms$ toward \lberr{336}{46}{38}{23}. The
error-bias-corrected amplitude is 384 \kms (to be compared with LP's
quoted 689 \kms).  A3381 is not in the Laine et al.\ sample, but is
the most discrepant cluster in the SMAC-LP distance comparison
performed in Paper IV (3.7 $\sigma$). While this cluster has a
somewhat larger internal FP scatter \citep{JorFraKja96}, its SMAC
peculiar velocity is not significant $1103\pm679 \kms$. In contrast,
its peculiar velocity according to LP is large and signficantly
different from zero ($-5900\pm2100 \kms$), so we suspect that for this
object LP are in error, rather than SMAC.  If we also remove A3381,
the LP bulk motion becomes $667\pm242\,\kms$ towards
\lberr{315}{48}{35}{26}.  This differs from no bulk flow, but only at
a marginal (95\%) confidence level. This is in better agreement (both
in direction and amplitude) with the SMAC bulk flow.  Along the
direction \lb{300}{10}, the motion of this sample is
$700\pm380\,\kms$. Note that because of the error ellipsoid is
triaxial it is possible that a flow solution fixed along a given
direction has a higher amplitude than that of the flow along the
direction of best fit.

There is therefore a strong hint that the original LP sample was
affected by dust. Although it is tempting to use the result from the
culled sample as a ``corrected'' LP bulk flow, it is important to
remember that there remain 25 galaxies in the BCG sample of LP which
are neither in Laine et al.\ sample nor have comparison peculiar
velocity data from SMAC.  It is likely that some ($\sim 25$\%) of
these will be strongly affected by dust.  Furthermore, it is also
possible that the LP motion is affected by other systematics, such as
the tendency for brighter BCGs to inhabit more X-ray-luminous clusters
\citep{HudEbe97}.  We conclude that the original LP result is in
marginal disagreement with SMAC, but this disagreement is no longer
significant once some BCGs with clear evidence of dust are removed
from the LP sample.

\subsubsection{Tully-Fisher distances to clusters}

\citet{MouAkeBot93} observed 38 TF clusters and fitted a bulk flow of
$811\pm138\,\kms$ towards \lb{332}{-15}. Along the ``concordance''
direction, this projects to $617\pm138\kms$. Excluding two outliers
(NGC 5419 and A3627), they obtained a flow of $559\pm107\,\kms$ toward
\lb{326}{-9} or $473\pm107\,\kms$ toward \lb{300}{10}.

\citet{Wil99b} found a bulk flow of $961\pm280\,\kms$ toward
\lb{266}{19} from 15 clusters using the Tully-Fisher relation.  The
amplitude and direction are in excellent agreement with SMAC bulk
flow, although the errors are large.  Along the ``concordance''
direction, the flow of the \citet{Wil99b} sample is $820\pm410\,\kms$.

In contrast, \citet{DalGioHay99b} found a bulk flow of only $75\pm92
\kms$ towards \lb{289}{25} (with large errors in the direction due to
the small amplitude of the bulk flow) from TF distances in 64 clusters
with a similar depth as the SMAC sample (but with different spatial
sampling).  The amplitude is consistent with zero.  If we fix the
direction to be \lb{300}{10}, our fits yield an amplitude of
$120\pm120$ \kms.  This is consistent with the SMAC flow in this
direction.

\subsubsection{EFAR} 

\citet{ColSagBur01} studied 50 clusters in two distinct regions of the
sky (Perseus-Pisces and Hercules--Corona-Borealis) via the FP
method. They found no significant bulk flow for the EFAR sample.

The geometry of the EFAR survey introduces several subtle problems.
First, there is substantial covariance between the monopole (FP
zero-point) and dipole terms, with a correlation coefficient of
$-0.77$ between the monopole term and the Galactic $y$-component of
the EFAR bulk flow.  Since the SMAC bulk flow is essentially along the
$y$ direction, this is clearly an important issue. Colless et al.\
fixed the zero-point of the EFAR FP relation by assuming that a subset
of their clusters were at rest. However, for an anisotropic sample
such as EFAR, this choice presupposes that there is no bulk flow.
They studied the effect of this choice and concluded that it was
smaller than their random errors.  While this is correct, it is not
negligible and should be included in the total errors. Second,
irrespective of the monopole covariance, the effects of sampling are
particularly severe for the EFAR sample, which covers two specific
regions on the sky.

We have fit the EFAR peculiar velocities to flow models including both
a bulk flow and free zero-point.  Note that, because we are fitting a
flow model to distances a posteriori (Method I of \citealt{StrWil95}),
we apply homogeneous Malmquist bias corrections to the EFAR cluster
distances. We also allow for distance errors due to errors in the
extinctions, in the same way as for the SMAC data. Note, however, that
the EFAR errors used in this analysis do not include the unpublished
systematic errors due to matching the velocity dispersions systems.
We expect these to be at a level at least that found for the SMAC
sample.

A fit of the EFAR data yields a bulk flow of $629\pm381\,\kms$ towards
\lberr{53}{41}{6}{25}.  If we force EFAR to have the same bulk flow as
the SMAC sample, we find $\Delta \chi^2 = 11$ for an increase of three
degrees of freedom. Thus the \emph{best fit} value of the SMAC bulk
flow is rejected at the 98.9\% CL.  This does not mean that EFAR
results are inconsistent with the SMAC results.  In the direction
\lb{300}{10}, the EFAR amplitude is $120\pm310\kms$, which is
consistent with the SMAC value $400\pm120\,\kms$.

\subsubsection{SNIa}

We have analyzed 98 SNe within 150 \hmpc\ from the compilation of
\citet{TonSchBar03}. Details of this analysis will be presented
elsewhere. In summary, we find that the sample as a whole has a bulk
flow of $410\pm75\kms$ towards \lb{286}{-12}. However the effective
depth of this sample is only $35 \hmpc$.  To test the coherence of the
flow, we have split the sample into two subsamples: SNIa-In ($0 < r <
60 \hmpc$) and SNIa-Out ($60 < r < 150 \hmpc$).  The SNIa-In sample
has a bulk flow of $376\pm81 \kms$ toward \lb{285}{-14}, which is not
surprising since this subsample spans a distance range where flows are
known to be high. The SNIa-Out sample has a bulk flow of $775\pm204
\kms$ toward \lberr{299}{17}{3}{13}.  The bulk flow of the outer
sample is significantly different from zero at the 99.8\% CL.  The
direction and amplitude of the SNIA-Out subsample are in good
agreement with the SMAC bulk flow.  

\subsubsection{Comparison of large-scale surveys}

\begin{figure}
\epsfxsize\columnwidth \epsfbox{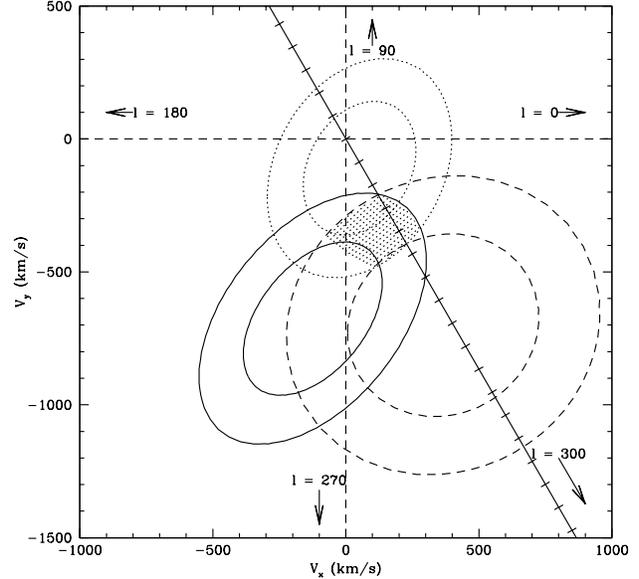}
\caption{ Galactic $x$ and $y$ components of the bulk flow and their
  68\% and 95\% confidence level error ellipsoids. We have plotted
  results for the large-scale sparse surveys SMAC (solid), Dale et al.
  (dotted) and SNIa with $60 \hmpc < d < 150 \hmpc$ (dashed).  These
  surveys all have very small components in the $z$ direction.
  Solutions that lie within the 95\% confidence region of all three
  surveys are hatched.  These three samples which differ the most in
  their flows and hence set the tightest bounds on the concordance
  region.  The concordance direction \lb{300}{10} is indicated by the
  solid diagonal line.  The EFAR and Willick results are not plotted
  as their errors are considerably larger. The results of Mould et
  al. are not plotted here but are also consistent with this
  concordance region. In contrast to the other surveys, the LP bulk
  flow vector does have a large $z$ component and so its motion is not
  well-represented by plotting ellipses in the $x$-$y$
  plane.\label{fig:ellipses} }
\end{figure}

\figref{ellipses} compares the error-ellipsoids in the $V_x$-$V_y$
plane for the SMAC, SNIa-Out and Dale et al samples. Note that these
errors are due to peculiar velocity errors only. They do not include
the effects of sparse sampling which would tend to further increase
the errors.  These three surveys are the ones which, with the
exceptionof LP, are most discrepant given their bulk flows and
errors. Despite these differences there is a ``concordance'' region in
which the flow is 250-450 \kms.  The bulk flows of other surveys
mentioned in this section cover this concordance region and do not
further constrain it.

\subsection{Field surveys within approximately $60 \hmpc$}

The peculiar velocity field has also been mapped at higher density
within smaller volumes closer to the LG.  For these samples, it is
possible to obtain a bulk flow which is less contaminated by sampling
effects.

\begin{enumerate}
  
\item \citet{GioHayFre98} studied a large sample of TF field galaxies.
  They found a bulk flow of $200\pm65\,\kms$ towards \lb{295}{25}for
  the volume extending to 6500 \kms.

\item Within 50 \hmpc, the bulk flow of the Mark III sample
  \citep{WilCouFab97} is
  $305\pm110\,\kms$ toward \lb{313}{29} when volume-weighted
  \citep{DekEldKol99} via the POTENT method. The corresponding result
  for the SFI survey is $310\pm80\,\kms$ toward \lb{299}{29}.
  \citet{DekEldKol99} argue that this flow is generated from beyond 50
  \hmpc.

\item The Shellflow project \citep{CouWilStr00} studied TF galaxies in
  the range $4500 < cz < 7000\,\kms$ and found $V = 70^{+100}_{-70}
 \,\kms$, with a 95\% upper limit of $V < 300\,\kms$.  Due to the small
  amplitude of the flow in comparison with the errors, the direction
  is not well defined: \lberr{144}{140}{50}{79}.

\item The ENEAR sample of early-type galaxies within $cz < 7000\,\kms$
  has a bulk flow of $V =
  220\pm{60\,\mathrm{(random)}}\pm{50\,\mathrm{ (systematic) }}\,\kms$
  toward \lberr{304}{16}{25}{11} \citep{daCBerAlo00b}.

\item \citet{TonBlaAjh00} did not measure a bulk flow of their SBF
  sample. Instead, they fitted a combination of internal flows (from
  GA and Virgo) plus external dipole and quadrupole terms.  The
  external dipole is $\sim 150\,\kms$ toward \lb{294}{67}. The errors
  on this ampitude are difficult to estimate because of the large
  degeneracies with the GA infall, but are approximately 100 \kms.

\end{enumerate}

In summary, with the exception of Shellflow, the field surveys find a
consistent non-zero flow in a similar direction on the sky. Note,
however, that the three TF samples (Mark III, SFI, Shellflow) are not
independent.  A conservative estimate of the bulk flow within 60
\hmpc\ from field TF and FP data is $\sim 200\pm75\kms$.

\begin{figure}
  \epsfxsize\columnwidth \epsfbox{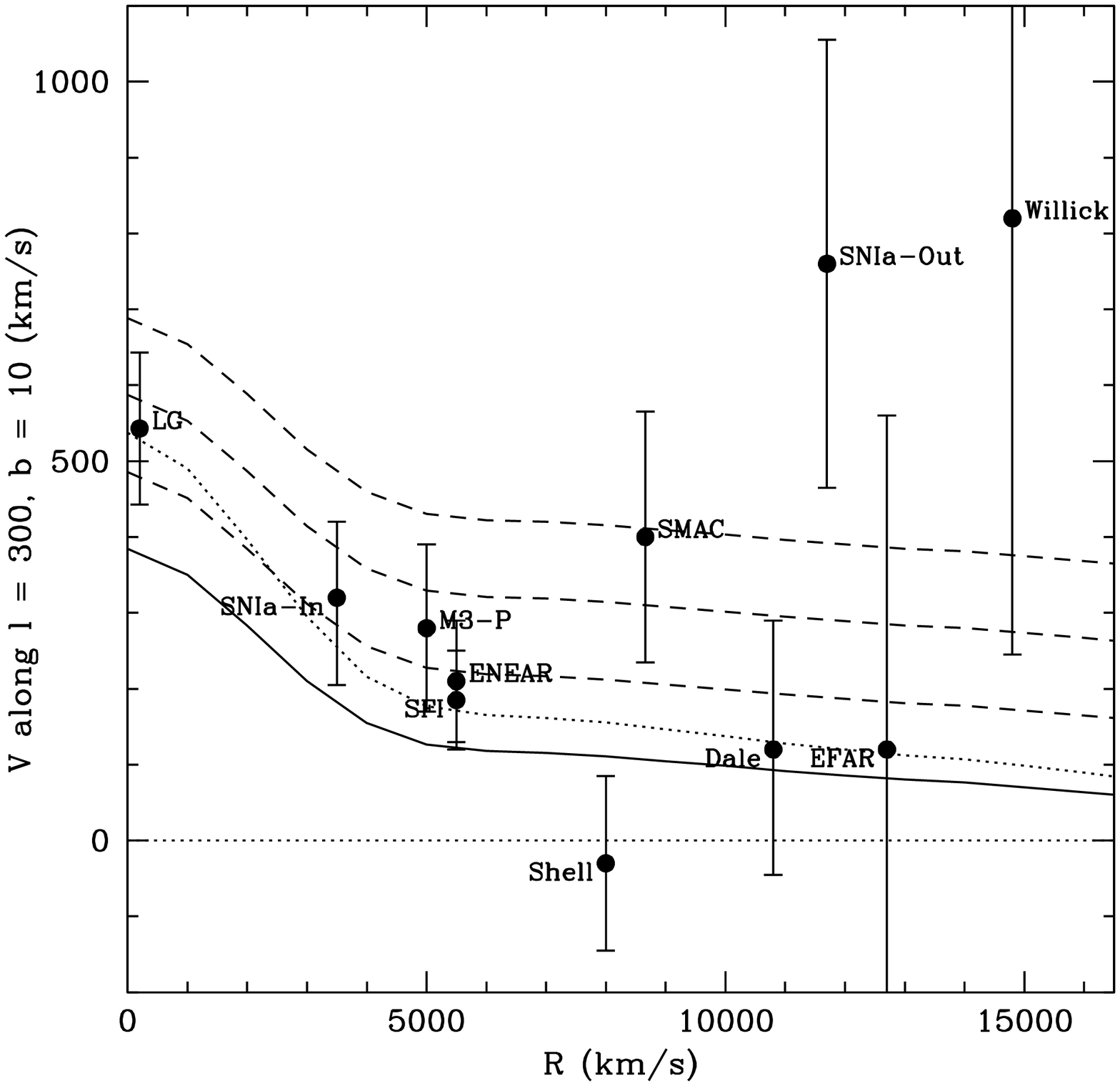}
\caption{ The bulk flow projected along the direction \lb{300}{10} as
  a function of depth, for the LG and various surveys: SNIa with $r <
  60 \hmpc$ (SNIa-In), \citet{DekEldKol99} analysis of the Mark III
  sample (M3-P), SFI, ENEAR, Shellflow, SMAC, distant ($60 \hmpc < r <
  150 \hmpc$) SNe (SNIa-out), EFAR and \citet{Wil99b}. Error bars are
  $1\sigma$ and do not include the effects of sparse sampling. Surveys
  are placed approximately at the depth corresponding to the top-hat
  radius of an idealized volume-weighted survey with the same
  effective depth.  This estimate is very crude: horizontal error bars
  on the depth are $\sim 10 \hmpc$ for the nearby samples and $\sim 20
  \hmpc$ for the distant ones. The solid curve shows the bulk flow
  predicted from the PSCz survey if $\beta\sbr{I} = 0.5$, whereas the
  dotted line shows the predicted bulk flow for $\beta\sbr{I} = 0.7$.
  The dashed lines show the PSCz bulk flow with $\vext$ of 100, 200
  and 300 \kms.
\label{fig:conc}
}
\end{figure}

\subsection{Towards a consistent model for large-scale flows}

The bulk flows of all surveys discussed above, projected into the
direction \lb{300}{10} are shown in \figref{conc}.  Surveys are
plotted as a function of their estimated depth. The lines indicate
\emph{IRAS} PSCz predictions for the bulk flow along this direction
for different values of $\beta\sbr{I}$ and $\vext$.  An eyeball fit
reveals that $\vext = 0$ is excluded by the distant samples, in
particular SMAC and the SNIa-Out sample.  Similarly, a bulk flow with
$V = 600 \kms$ is also excluded, in particular by \citet{DalGioHay99b}
and Shellflow.

A reasonable fit to all samples is possible if $\beta\sbr{I} = 0.5$,
producing a bulk flow component of $\sim 100$ \kms\ in this direction
at $R=100 \hmpc$ from sources in the PSCz survey, plus an additional
$\vext$ component from sources not included in PSCz.  Most of the
field peculiar surveys are consistent $\vext$ of approximately 125
\kms, with the exception of Shellflow which prefers $\vext \sim
0$. The deeper surveys prefer a higher value $\vext \sim 200 \kms$.
It is difficult, however, to quantify the effects of sparse sampling,
particularly on the latter surveys. Inspection of \figref{conc}
suggests that $\vext = 125 \kms$ is consistent with almost all of the
peculiar velocity surveys. There appear to be two exceptions:
Shellflow, which lies too low, and the SNIa-Out sample, which is too
high.  Since both of these samples have unusual geometry --- Shellflow
because it is a shell, and the SNIa-Out sample because it is sparse
--- it is likely that sampling errors will contribute significantly to
the error bars.

By adding $\vext = 125 \kms$ to the 100 \kms\ contribution from the
PSCz density field, one finds that 225 \kms, or a third of the LG's
motion arises from sources at large scales ($R > 100$ \hmpc).  Given
the systematic uncertainties and effects of sampling, we estimate the
error on this value is approximately 100 \kms.

\section{Conclusions}

We have examined in detail the $687\pm203$ \kms\ bulk flow of the SMAC
sample and find that it is robust to systematic errors.  We have shown
that most of the bulk flow is not generated by nearby sources such as
the Great Attractor but rather arises from structures at depths
greater than 100 \hmpc.  The Shapley Concentration is identified as
one likely source of the large scale flow but is unlikely to be
responsible for all of the SMAC flow.

When we compare the SMAC motion to the predictions of the PSCz survey
we find that $\beta = 0.39\pm0.17$, consistent with the
``concordance'' value $\beta \sim 0.5$.  However, the \emph{IRAS} PSCz
survey can only explain $50\pm20$\% of the amplitude of the SMAC flow.
This suggests that there are gravitational sources not well-mapped by
the PSCz survey.  Evidence from other redshift surveys suggests that
these sources may be located in the ZOA or in superclusters such as
Shapley which are undersampled by \emph{IRAS}.

Finally, we have shown that the SMAC survey is not inconsistent with
other sparsely-sampled surveys of the large scale velocity field, such
as \citet[EFAR]{DalGioHay99b,Wil99b} and the SNe sample of
\citet{TonSchBar03}. Taken together, all surveys suggest a large scale
flow approximately 225 \kms\ towards \lb{300}{10}.  Further analyses
of the effects of sparse sampling, and detailed comparisons with the
predictions of redshift surveys such as PSCz are needed to more
accuractely quantify this result.

The NOAO Fundamental Plane Survey \citep{SmiHudNel04} will measure the
peculiar velocities of 100 X-ray selected clusters.  The total number
of FP distances will be $\sim 4000$, about six times the number in
SMAC.  When results from the NFPS are compared to, for example,
predictions of the 2MASS redshift survey \citep{HucJarSkr03}, we
expect to identify the sources the LG's motion.

\section*{acknowledgments}

The data used in this project were obtained at the Isaac Newton Group
of telescopes at the Observatorio del Roque de los Muchachos, the
Anglo- Australian Observatory, and the Cerro Tololo Inter-American
Observatory. MJH acknowledges financial support from the NSERC and a
Premier's Research Excellence Award.

\end{document}